\documentclass[aps,prb,showpacs,superscriptaddress,twocolumn,10pt,floatfix]{revtex4-1}
\usepackage{graphicx}
\usepackage{amsmath}
\usepackage{amssymb}
\usepackage{epsfig}
\usepackage{physics}
\usepackage{multirow}
\usepackage{siunitx}
\newcommand{\ep}{\varepsilon}

\begin{document}

\title{Ballistic transport through irradiated graphene}

\author{Jonathan Atteia}
\email{jonathan.atteia@u-bordeaux.fr}
\affiliation{LOMA (UMR-5798), CNRS and Bordeaux University, F-33045 Talence, France}

\author{Jens H. Bardarson}
\email{jensba@pks.mpg.de}
\affiliation{Department of Physics, KTH Royal Institute of Technology, Stockholm, SE-106 91 Sweden}
\affiliation{Max-Planck-Institut f\"ur Physik komplexer Systeme, N\"othnitzer Str. 38, 01187 Dresden, Germany}

\author{J\'er\^ome Cayssol}
\email{jerome.cayssol@u-bordeaux.fr}
\affiliation{LOMA (UMR-5798), CNRS and Bordeaux University, F-33045 Talence, France}

\date{\today}
\begin{abstract}
The coherent charge transport through an illuminated graphene ribbon is studied as function of electronic doping, frequency and strength of the electromagnetic driving, for monochromatic circularly polarized light. We focus on the DC current carried by 2D bulk carriers which is dominant (over edge transport) for short and wide enough samples. Broad dips in conductance are predicted for one-photon and multi-photon resonances between the valence and conductance bands. The residual conductance can be associated with evanescent states and related to dynamical gaps in the Floquet quasi-energy spectrum.     
\end{abstract}

\maketitle

\section{Introduction}

Topological insulators\cite{Hasan2010,Qi2011} are a novel class of materials which have attracted a lot of attention in the last few years. These materials possess a bulk band gap, and due to a specific topological order of the bands, robust metallic states appear at their boundary. The topologically protected edge or surface states arise from the particular crystalline structure of the material, which is unfortunately not easy to engineer or manipulate. Recently, it has been predicted that topologically trivial materials can be turned into topological insulators by driving them with a time-periodic external perturbation\cite{Oka2009,Inoue2010,Lindner2011,Kitagawa2010,Rudner2013}. The formalism used to study those periodically driven systems is based on the Floquet theorem\cite{Shirley1965,Sambe1973}, which is the analog of the Bloch theorem for periodic potentials in time. These so-called  Floquet topological insulators\cite{Lindner2011,Cayssol2013} (FTI's) have been classified using new types of topological invariants\cite{Kitagawa2010,Gomez2013,Rudner2013}. 

An important candidate for the realization of a condensed matter FTI is irradiated graphene. Two distinctive situations may be encountered depending on whether the photon energy $\hbar \omega$ is larger or smaller than the bandwidth of the crystal. In both cases, a gap opens at quasi-energy $\ep=0$ realizing a Haldane topological insulator with photo-induced chiral edge states\cite{Oka2009,Kitagawa2010,Gu2011,Lindner2011}. In the latter case, an additional dynamical gap at $\ep=\hbar\omega/2$ may also open due to one-photon resonance between the two bands of graphene. Photoinduced chiral edge states appear in the gaps, thereby linking the distinct Floquet bands\cite{Perez-Piskunow2014,Usaj2014,Kundu2014,Perez-Piskunow2015,Quelle2016}. The topological nature of these edge states and their connexion to transport properties (quantized Hall conductance) has been extensively studied \cite{Gu2011,Kitagawa2011,Foa2014,Kundu2014,Fruchart2016} mostly in the absence of dissipation. Dissipation, occurring by coupling to phonons, leads to non-trivial occupation probabilities which usually gives rise to a non-quantized Hall conductance\cite{Mitra2014,Mitra2015a,Mitra2015b,Mitra2016,Seetharam2015}. The Floquet-Bloch states have been observed by irradiating the surface states of a 3D topological insulator with circularly polarized mid-infrared light and measuring the Floquet spectrum through time resolved ARPES pump-probe experiments\cite{Wang2013}. Note that FTIs have also been realized in cold atom systems\cite{Jotzu2014} and in photonic crystals \cite{Rechtsman2013}.

Our main motivation is to suggest the use of DC transport measurements to investigate the features of the Floquet quasi-energy spectrum in coherent driven electronic systems. We therefore compute the conductance of graphene transistors driven by an electromagnetic  wave, typically in the terahertz (THz) or infrared (IR) range. Indeed photon energies in the range $10-100$ meV can be matched by chemical potential variations in typical graphene samples. Recently, photon-assisted shot-noise has been studied experimentally in coherent diffusive graphene samples irradiated in the THz range\cite{Parmentier2016}.  

In this paper, we consider a graphene based field effect transistor (gFET), whose conduction channel is irradiated by a circularly polarized electromagnetic wave at normal incidence (Fig. \ref{fig:transistor}). The carrier density (in absence of irradiation) can be tuned using a DC electrostatic backgate. The source and drain leads are heavily doped and are not irradiated. The two-terminal differential conductance is studied as a function of the irradiation strength, chemical potential $\mu$ in the central region, and photon energy $\hbar \omega$ for various lengths $L$ and widths $W$ of the graphene ribbon (Fig.\ref{fig:transistor}). We consider the ballistic regime which has been reached experimentally in high-mobility suspended or encapsulated graphene samples\cite{Calado2014,Banszerus2016,Wang2013One-Dimensional}. The photon energies are in the range $10-100$ meV, therefore much smaller than the electronic band width of graphene. In this regime, during their time of flight across the transistor, the electrons are dressed coherently by the photons. The Floquet-Landauer-B\"{u}ttiker scattering theory allows to evaluate the source-drain conductance of the gFET. This approach implicitly assumes that the dissipation takes place only in the leads, and not in the central irradiated region. This assumption puts a restriction in the length $L$ of our sample. Due to the low electron-phonon scattering\cite{Balandin2011} in graphene, especially below the optical phonon energies ($200$ meV), we expect that dissipation can be neglected in sub-microns samples at low enough temperatures. Experimentally, the conductance in a coherent ballistic periodically-driven conductor has not been investigated yet for all $\mu$, and most theoretical studies focused so far on the case $\mu=0$\cite{Oka2009,Kitagawa2011,Lindner2011,Gu2011}. For doped graphene samples, the conduction by bulk states is always important and often exceeds the edge state contribution. Therefore this paper mainly focuses on the bulk conduction through 2D evanescent and/or propagating electronic states. Finally the response of a graphene-based transistor to electromagnetic driving remains a crucial issue for optoelectronic applications\cite{Mak2012}. 


The main results are the following. First, the conductance can be strongly suppressed in broad domains of chemical potential (or equivalently carrier density). The qualitative picture is that time-dependent driving opens gaps in the quasi-energy spectrum, inducing evanescent states characterized by finite penetration lengths. Hence, the conductance-chemical potential curves depend drastically on the length of the central conducting channel. Second, at neutrality, the minimal conductance exhibits non-monotonic variations as function of the irradiation strength. This is due to the fact that at the Dirac point, the conductance is exclusively carried by evanescent states even in the non irradiated case. Therefore the irradiation modifies the structure of evanescent states. 

The paper is organized as follows. In Sec. II, the Floquet quasi-energy spectrum of irradiated graphene is reviewed  \cite{Zhou2011,Oka2009,Usaj2014,Perez-Piskunow2015,Gu2011}. In Sec III, the scattering problem is formulated and the expression for the differential conductance is derived within the Floquet-Landauer-B\"{u}ttiker formalism. Sec. IV is devoted to the dependence of the conductance as function of $\mu$ and irradiation, while we focus on the case of undoped graphene ($\mu=0$) in Sec. V. Suitable parameters for experimental realizations are discussed in Sec. VI.

\section{Floquet spectrum in graphene}

In this section, we review the Floquet quasi-energy spectrum of an infinite graphene sheet irradiated by a circularly polarized electromagnetic wave \cite{Zhou2011,Oka2009,Usaj2014,Perez-Piskunow2015,Gu2011}. The purpose of this section is to set the stage for the transport calculations (sections III, IV and V) and make the paper self-contained.

\subsection{Geometry of the system and Hamiltonian}

The graphene sheet is irradiated by a circularly polarized radiation of frequency $\omega=2\pi/T$. Photon energies $\hbar\omega$ are much smaller than the bandwidth of graphene, which allows us to use the low-energy model for graphene. The massless Dirac fermions come in two flavors located around the two corners (valleys) of the Brillouin zone. Moreover, we consider ballistic graphene and only vertical (namely conserving momentum) electronic transitions (absorption or emission of photons) are allowed. Therefore inter-valley scattering can be neglected, and the decoupled ($K$ and $K'$) valleys can be studied separately.

\begin{figure}[h!]
	\includegraphics[width=8cm]{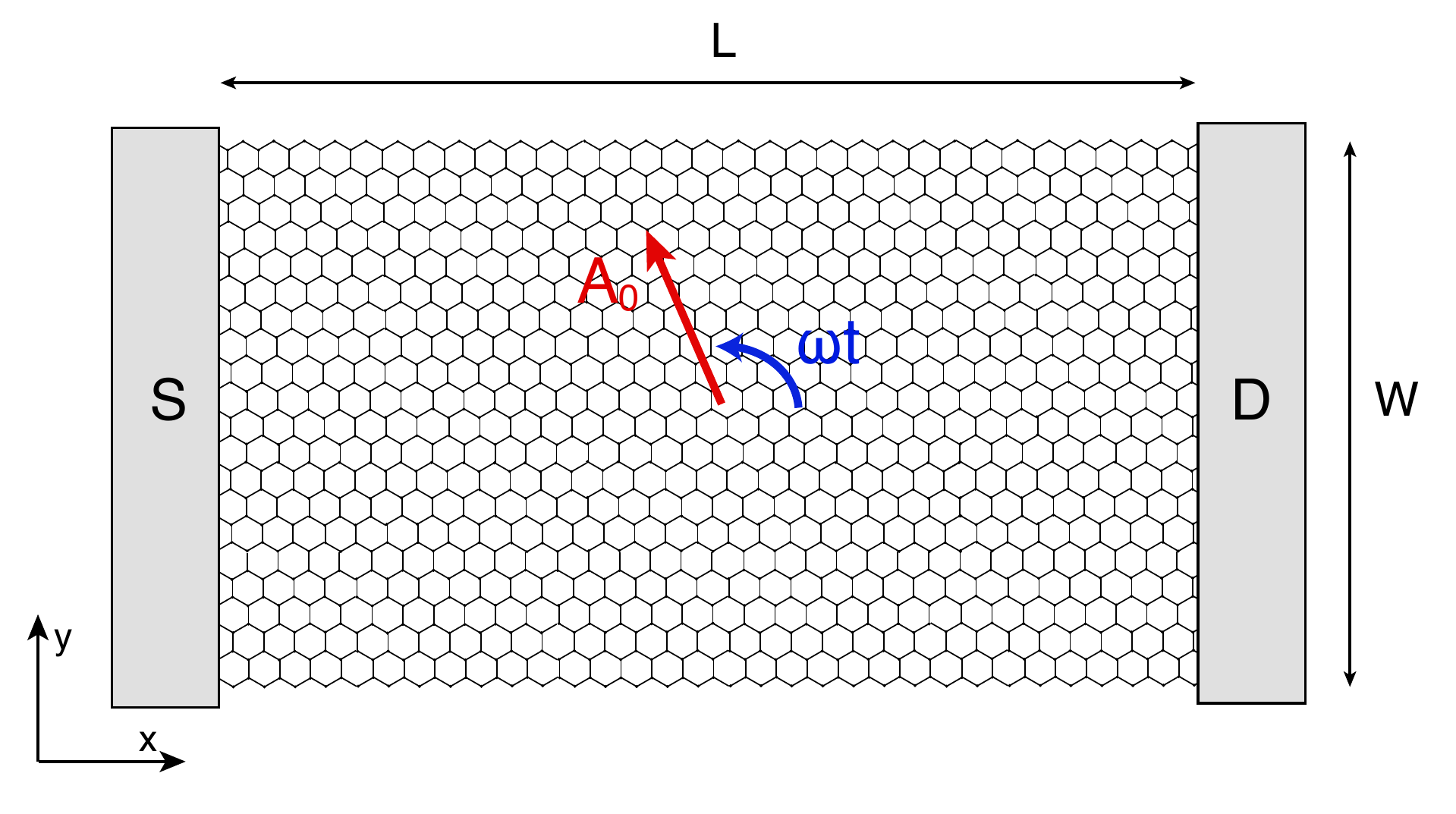} \\
	\caption{Field effect transistor consisting in a rectangular graphene ribbon contacted between source (S) and drain (D) leads. A remote electrostatic gate (not represented) allows to tune the chemical potential (and carrier density) in the central graphene region. A 
	circularly polarized electromagnetic wave is shone at normal incidence, corresponding to an in-plane rotation vector potential $\mathbf{A}(t)=A_0(\cos\omega t,\sin\omega t)$.}
	\label{fig:transistor}
\end{figure}

The time-dependent Hamiltonian describing the Dirac fermions in valley $\xi$ is written as :
\begin{equation}
H^\xi (t)=H_0^\xi  + V^\xi (t) \, ,
\label{eq:Hamiltonian} 
\end{equation}
$\xi=\pm 1$ being the valley index. 

The time-independent Hamiltonian $H_0^\xi$, corresponding to non irradiated graphene, reads\cite{Geim2009,Goerbig2011} : 
\begin{equation}
H_0^\xi = v (\xi\sigma_x p_x + \sigma_y p_y)- \mu ,
\label{eq:non_irrad_Hamiltonian}
\end{equation}
where $\sigma_x$ and $\sigma_y$ are the Pauli matrices associated with the sublattice isospin, $(p_x,p_y)=-i\hbar (\partial_x ,\partial_y )$ are the components of the wave-vector operator $\mathbf{p}$, $v$ is the Fermi velocity, and $\mu$ is the chemical potential. 

The vector potential $\mathbf{A}(t)=A_0(\cos\omega t,\sin\omega t)$ couples to the electric charge via the Peierls substitution $\mathbf{p}\rightarrow\mathbf{p}-q\mathbf{A}(t)$, where $q=-e$ is the electron charge. The Hamiltonian for the coupling between electrons and the electromagnetic field can therefore be expressed as:
\begin{align}
V^\xi (t)=(V_{1}^\xi  e^{i \omega t} + V_{-1}^\xi  e^{ -i \omega t}),
\label{potential}
\end{align}
where the matrices $V_1^\xi $ and $V_{-1}^\xi $ read:
\begin{equation}
V_{1}^\xi=\frac{evA_0}{2}(\xi\sigma_x - i\sigma_y), \ \ V_{-1}^\xi=\frac{evA_0}{2}(\xi\sigma_x + i\sigma_y).
\label{V1}
\end{equation}

An important dimensionless parameter is 
\begin{equation}
\beta=\frac{evA_0}{\hbar\omega} \, ,
\label{eq:beta} 
\end{equation}
which quantifies the electromagnetic driving strength. Following a simple quasi-classical argument, $evA_0$ is simply the energy gained by an electron traveling at speed $v$ in an electric field $E_0 = \omega A_0$ during a period of the electromagnetic wave ($1/\omega$), while $\hbar \omega$ is the minimal energy quantum which can be absorbed by the electron. The typical electric field can be expressed as:
\begin{equation}
E_0 = \beta \frac{\hbar\omega}{e l_\omega} \, ,
\label{eq:electric_field} 
\end{equation}
where $l_\omega = v / \omega$ is the distance travelled by the electron during a period of the electromagnetic field (divided by $2 \pi$).

\subsection{Floquet spectrum}

In this section, we assume translational invariance along the $x$ and $y$ axes and seek for solutions of the form $\Psi(t) e^{i k_x x+i k_y y}$. According to the Floquet theorem, the time-dependent part of the wave functions can be written as: 
\begin{equation}
\Psi(t) = \Phi_\ep(t) e^{- i \ep t/\hbar}  \,  ,
\label{Floq_state} 
\end{equation}
where $\Phi_\ep(t)$ is a $T$-periodic function and $\ep$ is the corresponding quasi-energy. The quasi-energy $\ep$ is defined modulo $\hbar \omega$, thereby making it possible to redefine $\ep$ such that it always belongs to the interval $[-\hbar\omega/2,\hbar\omega/2)$.

The wave function obeys the Schr\"{o}dinger equation
\begin{equation}
  i\hbar\partial_t   \Psi(t)  = H^\xi (t)   \Psi(t).
 \label{eq:Schro_eq}
\end{equation}
After injecting Eq.(\ref{Floq_state}) in the Schr\"{o}dinger equation, the $T$-periodic function $\Phi_\ep(t)$ appears to be an eigenstate of the Floquet Hamiltonian $H_F(t)=H(t)-i\partial t$ with eigenvalue $\ep$, namely:
\begin{equation}
 \left(H^\xi (t) - i\hbar\partial_t   \right)  \Phi_\ep(t)  = \ep   \Phi_\ep(t).
 \label{eq:Floquet_eq}
\end{equation}
Using the following Fourier expansion for $\Phi(t)$:
\begin{equation}
 \Phi_\ep(t) = \sum_{m \in \mathbb{Z}} \Phi_m e^{i m \omega t}  \, ,  
\label{Floq_fourier} 
\end{equation}
allows to transform the time-dependent differential equation Eq.(\ref{eq:Floquet_eq}) into a time-independent eigenstate problem\cite{Sambe1973,Shirley1965}:
\begin{equation}
 \sum_n H_{F,mn}^\xi \Phi_n = \sum_n (H^\xi_{0F,mn} +V_{F,mn}^\xi)  \Phi_n =\ep  \Phi_m \, ,
 \label{eq:FloquetFourier}
\end{equation}
where the quasi-energy index $\ep$ is omitted in all Fourier components $\Phi_n$, and $(m,n) \in  \mathbb{Z}^2$. The Floquet Hamiltonian $H^\xi_{0F,mn}$, defined as
\begin{equation}
\label{Floq_mat1}
H_{0F,mn}^\xi=(\hbar v(\xi\sigma_x k_x  + k_y \sigma_y) - \mu  +  m \hbar\omega )  \delta_{mn}, \\
\end{equation}
is diagonal in the Floquet basis. In contrast, the driving Hamiltonian $V_{F,mn}^\xi$,   
\begin{equation}
\label{Floq_mat2}
V_{F,mn}^\xi =V_{1}^\xi \delta_{m,n-1} + V_{-1}^\xi \delta_{m,n+1},
\end{equation}
couples distinct Fourier components $\Phi_n$. Matrices $V_{1}^\xi$ and $V_{-1}^\xi$ are defined by Eq.(\ref{V1}).

In this Floquet-Fourier representation, the wave function is an infinite vector containing all the different Fourier harmonics of the wave function. Since the Floquet matrix is infinite, we need to set a cut-off $N$ in the number of Floquet replicas considered. The size of the matrix is thus $2(2N+1)\times2(2N+1)$ (taking into account the sublattice isospin index). Afterwards, it will be necessary to check the stability of the results by increasing $N$.

\begin{figure}[h!]
\includegraphics[width=8cm]{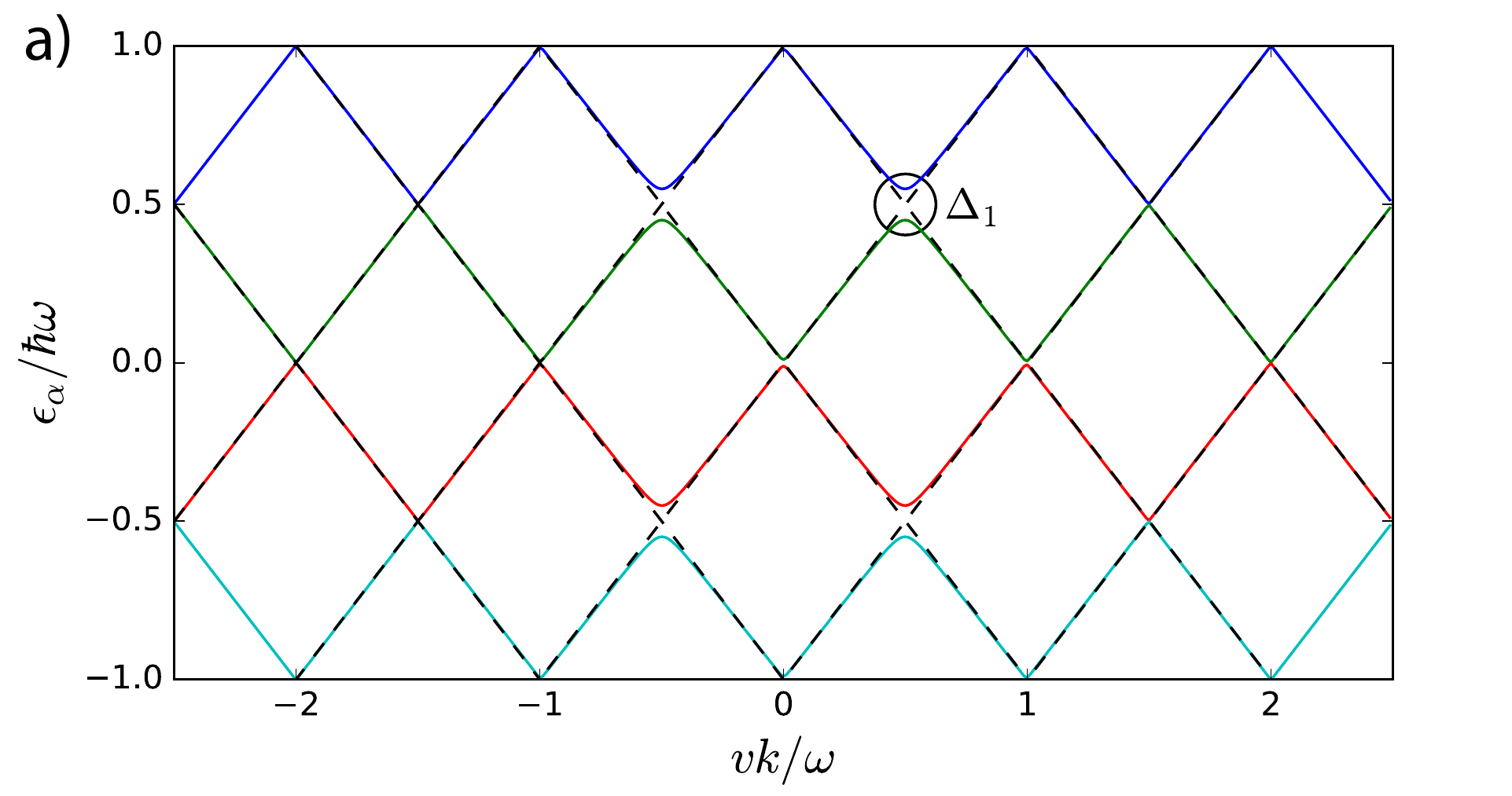}

\includegraphics[width=8cm]{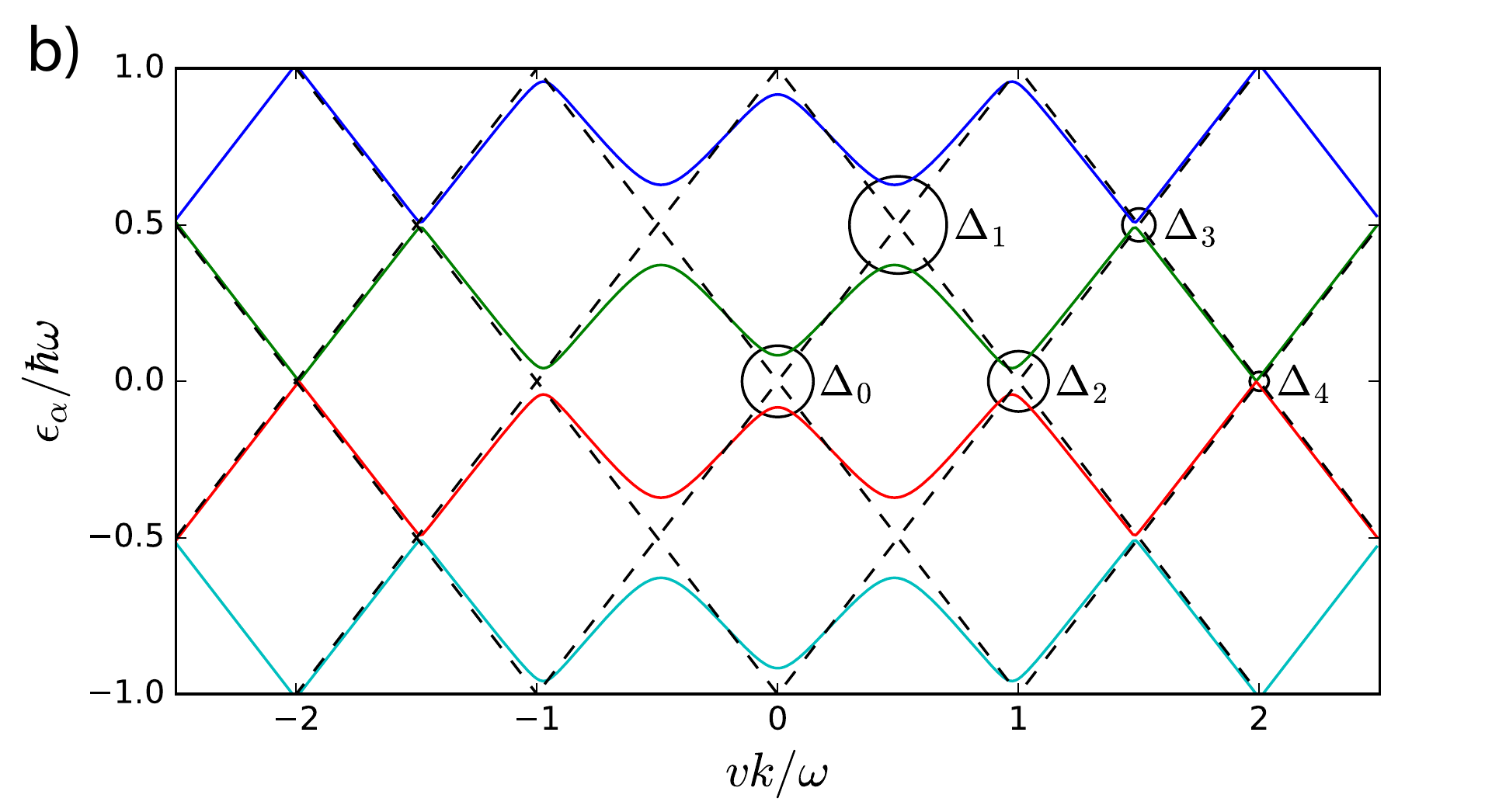}

\includegraphics[width=8cm]{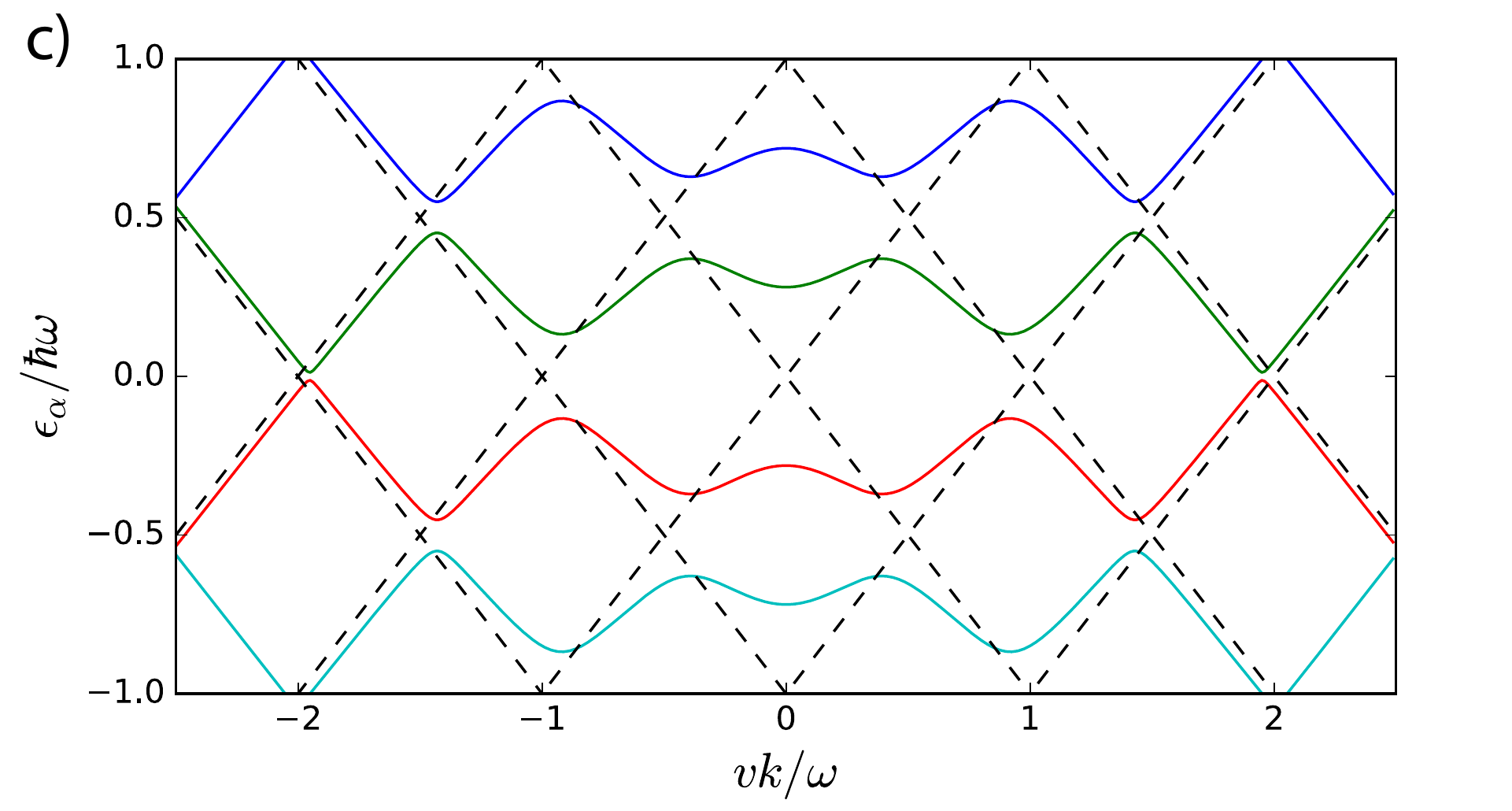}
\caption{ Dispersion relation $\varepsilon_\alpha(k)$ ($k=|\mathbf{k}|$) of an infinite graphene sheet irradiated by an electromagnetic wave at various driving strengths : a) $\beta=0.1$, b) $\beta=0.3$ and c) $\beta=0.6$ using $N=3$ Floquet replicas. The dashed lines correspond to the non-irradiated case in the Floquet representation with dispersion $\ep=\pm\hbar vk+n\hbar\omega$. The circles on a) and b) indicate the non-equivalent gaps $\Delta_m$, located at momenta $k=m\omega/2v$, and at quasi-energies $\ep=0$ and $\ep=\hbar\omega/2$, alternatively. }
\label{fig:spectrum}
\end{figure}

Once the cut-off is made, the Floquet matrix $H_{F}^\xi$ can be easily diagonalized numerically to obtain the dispersion relation. Due to rotational invariance around $\mathbf{k}=0$, the spectrum only depends on the norm $k=|\mathbf{k}|$ of the wave vector. The spectrum consists of $2(2N+1)$ bands having the dispersion relation $\varepsilon_\alpha(k)$ for $\alpha\in[1,2(2N+1)]$. The Floquet bands centered around $\varepsilon=0$ are plotted as functions of $k$ in Fig. \ref{fig:spectrum} for various driving strengths, using seven Fourier components ($N=3$).

For weak driving, $\beta=0.1$ [Fig \ref{fig:spectrum}.a)], the bands corresponding to the irradiated case (solid curves) follow closely the non-irradiated ones (dashed curves) except around avoided crossings $\Delta_1$ located at $k=\pm \omega/2v$ and centered around $\ep=\pm \hbar \omega/2$. Upon increasing the driving, more gaps open at momenta $k_m=\pm m\omega/2v$, with $m$ integer [Fig \ref{fig:spectrum}.b)]. These gaps correspond to the anti-crossing of two Dirac bands dressed with $n$ and $n'$ photons such that $|n-n'|=m$. For example, at the gap $\Delta_3$ in Fig. \ref{fig:spectrum}.b), the electron is in a coherent superposition of a valence band electron dressed with $2$ photons (having absorbed two quanta of the electromagnetic field) and a conduction band electron dressed with $-1$ photon (meaning having emitted one quantum of the field).

For strong driving, multi-photon processes are more likely, leading to a strong modification of the quasi-energy spectrum [Fig. \ref{fig:spectrum}.c)]. 

In the next sections, we study with more detail the gap structure as a function of $\beta$. The spectrum is periodic in energy (with period $\omega$), so we restrict ourselves to the study of the reduced bands where $\ep\in[-\hbar\omega/2,\hbar\omega/2]$. The gaps are located at different momenta because of the folding of the dispersion relation in the reduced Floquet zone $[-\hbar\omega/2,\hbar\omega/2]$. In this representation, there are two non-equivalent set of gaps centered at $\ep=0$ and at $\ep=\hbar\omega/2$ [Fig. \ref{fig:spectrum}.b)]. The gaps of even order ($\Delta_0$,$\Delta_2$,$\Delta_4$...) are located in the center of the Floquet zone ($\ep=0$, Sec. \ref{sec:gaps0}) and the even order ones ($\Delta_1$,$\Delta_3$...) are located at the edge  ($\ep=\pm\hbar\omega/2$) (Fig. \ref{fig:spectrum}.b)).

\subsection{Photoinduced gaps vs $\beta$}

\label{sec:gaps0}

\begin{figure}[h!]
\includegraphics[width=8cm]{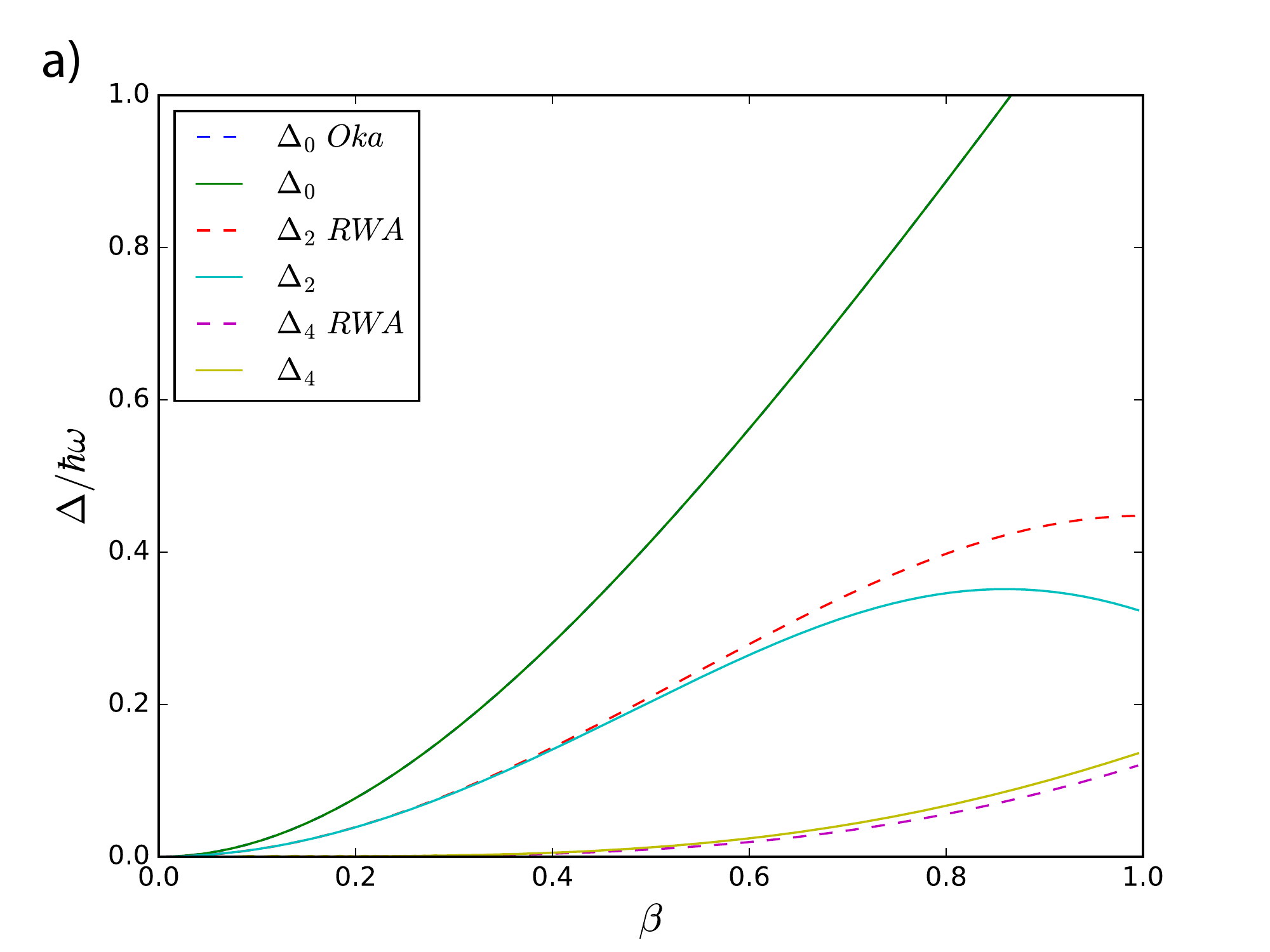} \\
\includegraphics[width=8cm]{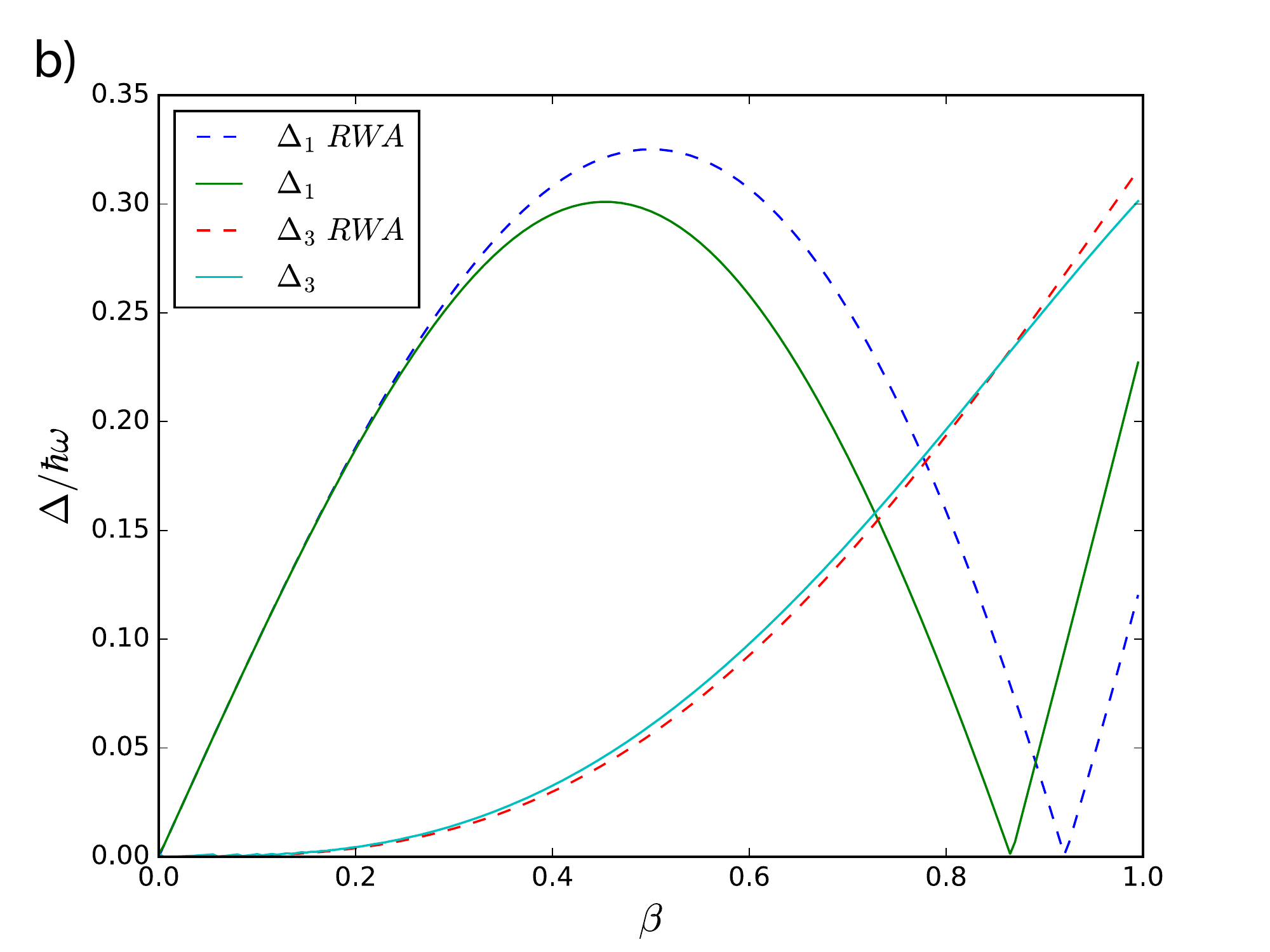}
\caption{Size of the gaps at a) $\mu=0$ and b) $\mu=\hbar\omega/2$ as a function of the driving strength $\beta$ calculated using the numerical Floquet method (solid lines) and the rotating wave approximation (RWA), namely Eqs. (\ref{eq:gap0}) and (\ref{eq:gapm}) (dashed curves). The numerical calculations (solid lines) agrees with RWA estimations for weak driving. The gap $\Delta_1$ closes while the gap $\Delta_0$ tends to $\hbar\omega$ for $\beta=\sqrt{3}/2$.}
\label{fig:gaps0}
\end{figure}

The Floquet spectrum has a very rich gap structure\cite{Perez-Piskunow2015} that depends strongly on the driving strength $\beta$. Understanding this structure is crucial in order to study the transport properties of irradiated graphene (see Sec \ref{sec:doped} and \ref{sec:undoped}). The gap $\Delta_0$ and the gaps $\Delta_m$ with $m\geqslant1$ have different origins. Multiphoton processes of order $m$ induce the gaps $\Delta_m$. For $m \geqslant 1$, the gap amplitudes have been derived within the generalized rotating wave approximation (RWA) in Ref. \onlinecite{Zhou2011}:
\begin{equation}
\frac{\Delta_m}{\hbar\omega}=\beta|J_{m+1}(2\beta)-J_{m-1}(2\beta)|\approx\frac{\beta^m}{(m-1)!} \, ,
\label{eq:gapm}
\end{equation}
where $J_m$ is the $m$-th order Bessel function, and the last approximation is only valid for $\beta\ll 1$. For weak driving, the size of the gap $\Delta_m$ ($m\geqslant 1$) is strongly reduced upon increasing $m$. Increasing the driving strength $\beta$, the gaps evolve in a non-monotonic way and can also close. In this section, we will compare analytical expressions of the gap size obtained within the rotating wave approximation scheme and the results that we get from diagonalizing in Floquet space (Fig. \ref{fig:gaps0}).  The RWA fits very well the numerics for low $\beta$, and describes qualitatively the size of the gaps for $\beta \lesssim 1$ (Fig. \ref{fig:gaps0}). 

At energy $\ep=0$, Oka and Aoki\cite{Oka2009} first discussed the central gap $\Delta_0$ at $k=0$ and found that :
\begin{equation}
\frac{\Delta_0}{\hbar\omega}=\sqrt{1+4\beta^2}-1\approx2\beta^2 ,
\label{eq:gap0}
\end{equation}
where the last approximation correspond to weak driving $\beta\ll1$. The quadratic dependence in $\beta$ for weak driving points out that this gap originates from a second order process in the coupling : the emission and absorption of a virtual photon. The analytic expression Eq. (\ref{eq:gap0}) fits perfectly our numerics. For weak driving, the evolution is quadratic in $\beta$, and as the driving increases, it becomes linear until the gap size reaches the size of the Floquet zone $\ep=\hbar\omega$ at $\beta=\sqrt{3}/2$. At this particular value of $\beta$, the gap $\Delta_1$ vanishes as shown in Fig. \ref{fig:gaps0}.b). This is a case of a band touching induced by strong electromagnetic driving.
 
For weak driving, the gap $\Delta_1$ is located at wave-vector $k=\pm\omega/2v$, and around energy $\ep=\hbar\omega/2$. This gap originates from the anti-crossing of the bands $n=1$ and $n'=0$, and therefore it is a first order process in $\beta$ (exchange of one photon). This explains why $\Delta_1$ is the largest gap for weak driving [Fig. \ref{fig:spectrum}.a) and b)]. The size of this gap first increases linearly with $\beta$, then it reaches a maximum value around $\beta\approx0.5$, and starts decreasing until it reaches zero [Fig. \ref{fig:gaps0}.b)]. We see that at the same value of $\beta$, the gap $\Delta_0$ reaches $\hbar\omega$ (Fig. \ref{fig:gaps0}), and as we have seen earlier, this happens for $\beta=\sqrt{3}/2\approx 0.87$. We conclude that this value corresponds to a band touching.

The gap $\Delta_2$ originates from the anti-crossing of two Dirac bands dressed with $n=1$ and $n'=-1$ photons. This is therefore a second order process in the coupling and is proportional to $\beta^2$ for weak driving. The idea is the same for the gap $\Delta_4$, it originates from the anti-crossing of bands dressed with $2$ and $-2$ photons, and is thus proportional to $\beta^4$ for weak driving.

\section{Formalism}

In this section, we present the formalism used to compute the 2-terminal conductance of a graphene-based transistor whose central 2D conducting channel is irradiated by circularly polarized light, at normal incidence. The transmission and reflection coefficients are evaluated within the Landauer-B\"{u}ttiker formalism extended to driven Floquet systems \cite{Moskalets2002}. Note that the electromagnetic field is uniform in the central gated graphene region, while the leads are not irradiated.

\subsection{Geometry and scattering problem}

We consider an irradiated rectangular ribbon of width $W$ and length $L$ connected to two non-irradiated leads (Fig. \ref{fig:transistor}). The Hamiltonian being time-dependent, the single electron energy is not conserved. However, due to the invariance under translations of time $T$, the quasi-energy $\ep$ is still a good quantum number. In the scattering problem, we consider electrons incoming at energy $\ep=0$, which corresponds to the Fermi level of the leads. The electrons propagating inside the irradiated region can emit or absorb one or several photon(s) of energy $\hbar\omega$. These inelastic scattering events are described by a set of transmission (reflection) coefficients $t_{n0}$ ($r_{n0}$) corresponding to the amplitude of being transmitted (reflected) with a final energy $\ep +n  \hbar \omega$.  

In order to determine those Floquet scattering coefficients, periodic boundary conditions are used along the $y$ direction. Hence the transverse wave-vector $k_y$ is quantized as :
\begin{equation}
k_y=\frac{2\pi n_y}{W} \, ,
\end{equation}
with $n_y$ a relative integer. Note that with such periodic boundary conditions, only bulk 2D states are investigated. In contrast to $k_y$ and $\mu$, the longitudinal wave-vector $k_x$ is not a good quantum number because the system is not translationally invariant along the $x$-axis. 

In the following, all the two-component wave functions are written as $\Phi(x,y,t)=\Phi(x,t) e^{i k_y y}$, and we will work with the $\Phi(x,t)$ wave functions ($k_y$ being omitted) corresponding to the effective 1D transport problem at a given $k_y$. In order to calculate the two-terminal conductance, we need to match the wave functions at the interfaces $x=0$ and $x=L$. 

\subsection{Irradiated ribbon}
\label{sec:irrad}

In the irradiated region ($0<x<L$), we need to find the Floquet eigenstates $\Phi_m$ and the longitudinal wave-vector $k_x$ corresponding to a given set $(\mu,k_y)$ (still having in mind the scattering problem for $\ep=0$ corresponding to an electron incident from the lead Fermi level). Multiplying Eq. (\ref{eq:FloquetFourier}) by $\xi\sigma_x$ and rearranging it allows to get both $k_x(\ep)$ and  $\Phi_m$ by solving the following equations:
\begin{equation}
\sum_n (K_{0F,mn}^\xi+K_{VF,mn}^\xi) \Phi_n =\hbar v k_x \Phi_m \,
\end{equation}
where one has defined an infinite matrix $K_{0F}$ with matrix elements:
\begin{equation}
K_{0F,mn}^\xi=\xi((\mu - m \hbar\omega  )\sigma_x  - i  \sigma_z\hbar v k_y) \delta_{mn}  \, ,\\
\end{equation}
and the infinite matrix $K_{VF}$ by the matrix elements:
\begin{equation}
K_{VF,mn}^\xi=-\xi\sigma_x V_{1}^\xi \delta_{m,n-1} -\xi\sigma_x V_{-1}^\xi \delta_{m,n+1} \,
\end{equation}
describing the effect of the electromagnetic field. 

Note that changing the valley index reverses the sign of $k_x$, and because the ribbon is space-inversion symmetric along the $x$ direction, the eigenvalues $k_x$ always come in pairs, therefore both valleys have the same contribution to the conductance. In practice, one can solve the scattering problem and compute the transport in a given valley, and multiply the single-valley result by $2$ to get the total conductance.

\begin{figure}
\includegraphics[width=8cm]{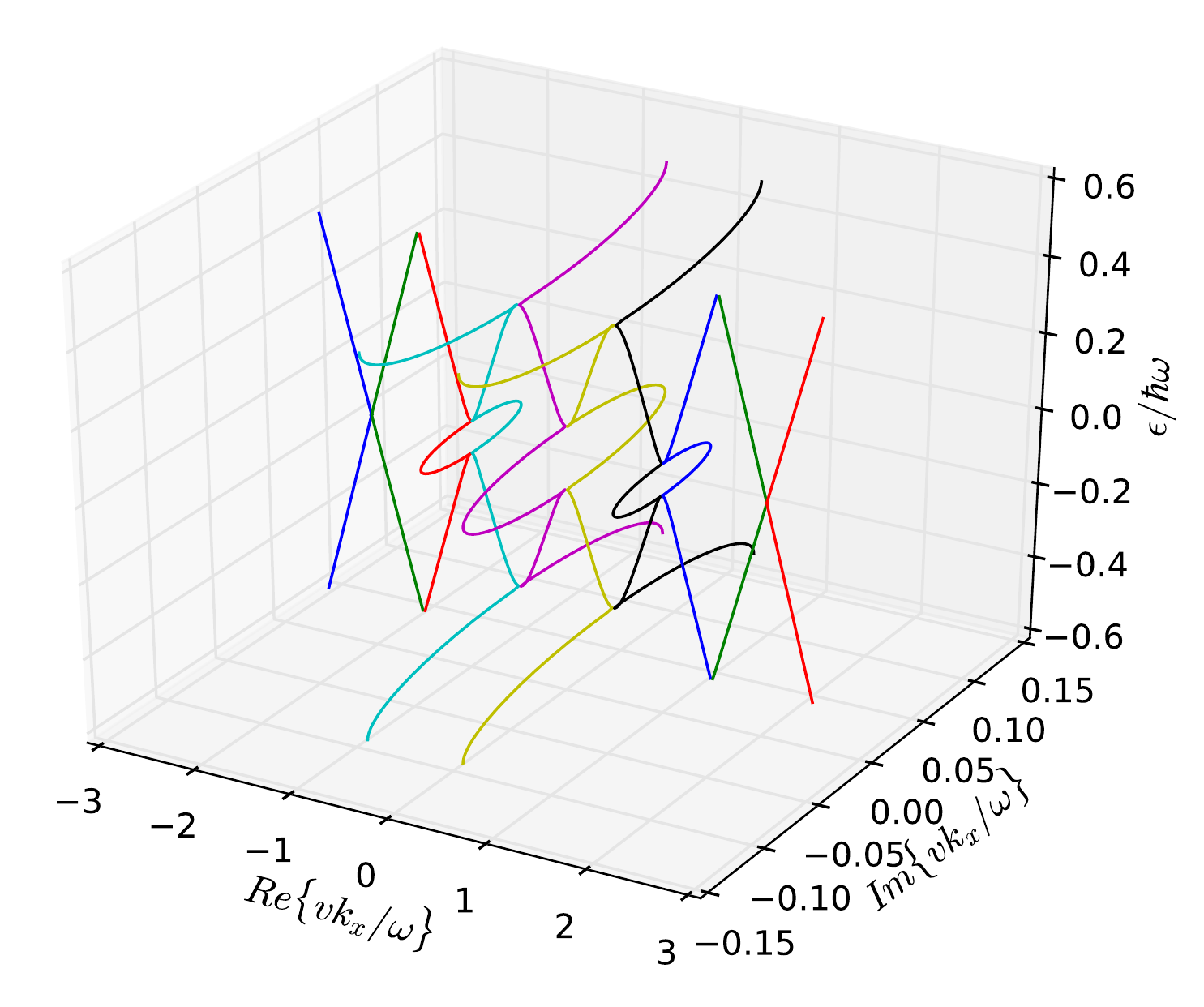}
\caption{Dispersion relation $k_x^\alpha(\ep)$ for $\beta=0.3$ in the reduced Floquet zone with the real and imaginary parts of the wave vector $k_x$ for $k_y=0$ and $N=2$.}
\label{fig:spec_imag}
\end{figure}

Besides, $K_F=K_{0F}+K_{VF}$ is not Hermitian so $k_x$ can have an imaginary part, which corresponds to an evanescent state (Fig. \ref{fig:spec_imag}). As we have seen before, we need to make a truncation in the number of Floquet side bands such that $m\in[-N,N]$.  After numerical diagonalization, we obtain $2(2N+1)$ eigenvalues $k_x^{\alpha}(\ep)$ and eigenvectors $\Phi^{\alpha}_m$ with $\alpha\in\{1,2(2N+1)\}$ labelling the different eigenmodes. The resulting 2-component wave function is written as a superposition of the $2(2N+1)$ eigenmodes labelled by index $\alpha$. The wave function of a given $\alpha$-eigenmode is expressed as: 

\begin{equation}
\Phi^{\alpha}(x,t) = e^{i k_x^{\alpha} x} \sum_{m=-N}^N  \Phi^{\alpha}_m    e^{- i m\omega t} , 
\end{equation}
which is still a two-component spinor (attached to one valley, here $\xi=1$).

When the Fermi level lies in the gap, $k_x^\alpha(\ep)$ has a non-zero imaginary part which signals an evanescent state. For a ribbon with a finite length, these states will allow the current to tunnel through the gapped central region. The penetration lengths $\xi$ are defined as the inverse of the imaginary part of the longitudinal wave-vector $k_x(\ep)$ inside the gap (Fig. \ref{fig:spec_imag}) :
\begin{equation}
\xi_\alpha=\frac{1}{|\Im\{ k_x^{\alpha}\}|}.
\end{equation}
For weak driving, $\beta\ll1$, it is possible to associate each eigenmode $(\alpha)$ with a given anti-crossing between uncoupled Floquet replicas. Moreover, there is a direct relation between the length $\xi$ and the size of the gaps :
\begin{equation}
\xi_m=\frac{\hbar v}{\Delta_m}.
\end{equation}
As the size of the gap increases, the imaginary part of $k_x(\ep)$ (where $\ep$ is in the gap) increases, thereby the characteristic length decreases. According to Eqs.~(\ref{eq:gap0}) and (\ref{eq:gapm}), and for $\beta \ll 1 $ the decay lengths are expressed as:
\begin{eqnarray}
\xi_0&=&\frac{l_\omega}{\sqrt{1+4\beta^2}-1}\approx\frac{l_\omega}{2\beta^2},
\label{eq:length_evan1}\\
\xi_m&\approx & \frac{(m-1)!}{\beta^m}  l_\omega \, , 
\label{eq:length_evan2}
 \end{eqnarray}
in terms of the length :
\begin{equation}
l_\omega=v/\omega\, ,
\end{equation}
corresponding to the distance travelled by an electron during one cycle of driving (divided by $2 \pi$). We compared the RWA method and the numerics and we see a good agreement between numerical diagonalisation and RWA even for $\beta\approx 1$.

\subsection{Leads}
To simplify the expression of the scattering states in the leads, we consider heavily $n$-doped leads, in the same manner as in Ref. \onlinecite{Beenakker2006} for the non-irradiated case. The chemical potential in the leads is $\mu_\infty>0$ for $x<0$ and $x>L$ such that $\mu_{\infty}\gg|\mu|$.

In order to account for electrons leaving the scattering region with energy $\ep+n\hbar\omega$, we use the Floquet theorem in the leads. However, because there is no driving, the Floquet replicas are decoupled.
In the leads, for an infinite system, Eq. (\ref{eq:Floquet_eq}) with $\ep=0$ becomes :
\begin{equation}
 \hbar v\left(  \sigma_x k_x  + \sigma_y k_y \right)  \Phi_m(x) =  (\mu_{\infty} - m \hbar\omega)   \Phi_m(x)
\end{equation}
For a given 2D wave-vector $(k_x,k_y)$, there is an infinite amount of plane wave solutions labelled by their Floquet index $m$ and their band index $s=\pm1$ :
\begin{equation}
\Phi_{m,s} (x)=\frac{1}{\sqrt{2}}\begin{pmatrix}
s  \\ 
e^{i \varphi}
\end{pmatrix} e^{i k_x x  }
\end{equation}
with $\mu_{\infty} =m\hbar\omega+ s\hbar v\sqrt{k_x^2+k_y^2}$ and $\cos \varphi = k_x / \mu_\infty$, $\varphi$ being the angle of incidence of the electron. Since $\mu_\infty \rightarrow \infty$, we have $\varphi \rightarrow 0, \pi$ and $s=+1$ in both source and drain leads. Finally, the spinor in the leads is independent of $m$ and we obtain one left $(+)$ and one right $(-)$ going solution :
\begin{equation}
\Phi_m^\pm (x)
=\Phi^\pm e^{i k_x^\pm x}  \, ,
\label{eq:wave_leads}
\end{equation}
with $k_x^\pm = \pm \mu_\infty$, and
\begin{equation}
\Phi^\pm =\frac{1}{\sqrt{2}}\begin{pmatrix}
1  \\ 
\pm  1
\end{pmatrix} 
  \, .
\label{eq:wave_leads1}
\end{equation}

\subsection{Matching at the interfaces}

Using the expression of the eigenstates in the different regions, it is possible to construct the scattering states. We consider an incoming wave from the left lead at the quasi-energy $\ep=0$ and Floquet index $n=0$. In the left lead, there will be $2N+1$ waves reflected at energies $\ep=n\hbar\omega$ with amplitude $r_{n0}$. The wave function in the left lead is then :
\begin{equation}
\Psi^L(x,t)  =  \left( \Phi^+ e^{i k_x^+ x} + \sum_{n=-N}^{N}  r_{n0}  \Phi^-  e^{i k_x^- x} e^{i n \omega  t}  \right)  e^{- i \ep  t} .
\end{equation}

In the irradiated region, the wave function is a superposition of the eigenstates $\Phi^\alpha$ with amplitudes $a_\alpha$ :
\begin{align}
\Psi^I(x,t)  &= \sum_\alpha   a_{\alpha}  \Phi^{\alpha}(x,t) e^{- i \ep   t} \nonumber  \\
&= \left( \sum_{\alpha} a_{\alpha} e^{i k_x^{\alpha} x}  \sum_{n=-N}^{N}    \Phi^{\alpha}_n   e^{- i n \omega t} \right) e^{- i \ep   t}.
\end{align}

In the right lead, only the right going states with amplitude $t_{n0}$ on the $n$'s replica are chosen :
\begin{equation}
\Psi^R(x,t)  =  \left(\sum_{n=-N}^{N}  t_{n0}  \Phi_n^+  e^{i k_x^+(x-L)} e^{i n \omega  t}  \right) e^{- i \ep  t}. 
\end{equation}

\

To calculate the reflected and transmitted amplitudes, we need to match the wave functions at the interfaces. The boundary condition at $x=0$ is $\Psi^L(x=0^-,t)=\Psi^I(x=0^+,t)$: 
\begin{equation}
\Phi^+ + \sum_{n=-N}^{N}   r_{n0}  \Phi_n^- e^{- i n \omega  t}  =  \sum_{\alpha} \sum_{n=-N}^N   a_{\alpha}   \Phi^{\alpha}_n   e^{- i n \omega t}  , 
\end{equation}
and at $x=L$, the condition is $\Psi^I(x=L^-,y,t)=\Psi^R(x=L^+,t)$ so :
\begin{equation}
\sum_{\alpha} \sum_{n=-N}^N   a_{\alpha} \Phi^{\alpha}_n e^{i k_x^{\alpha} L} e^{-i n \omega t} = \sum_{n=-N}^{N}   t_{n0}  \Phi_n^+ e^{- i n \omega  t}  \,. 
\end{equation}

These boundary conditions, at $x=0$ and $x=L$, must be valid at any time $t$, so we can project them on the different Fourier harmonics :
\begin{eqnarray}
\Phi^+\delta_{n0}  +    r_{n0} \Phi^- =  \sum_{\alpha} a_{\alpha}   \Phi^{\alpha}_n    , \\
\sum_{\alpha} a_{\alpha}  \Phi^{\alpha}_n e^{i k_x^{\alpha} L} = t_{n0} \Phi^+  .
\end{eqnarray}
Finally, on one hand, the number of unknown scattering parameters is $4(2N+1)$ since we have $(2N+1)$ reflection coefficients $ r_{n0}$, $(2N+1)$ transmission coefficients $t_{n0}$, and $2(2N+1)$ coefficients $a_\alpha$ to determine. On the other hand, each matching condition represents $2N+1$ spinor relations, hence there is a total of $4(2N+1)$ linear relations between those coefficients. 
 
The number of Floquet replicas $N$ defines the cut-off used to calculate the eigenstates and eigenvectors of the system, and therefore to obtain the transmission coefficients. We have systematically checked the convergence of our results with respect to $N$. 

\subsection{Conductance formula}

To calculate the conductance of the sample, we use the scattering theory extended to Floquet systems \cite{Moskalets2002}. The scattering matrix element $t_{n0,k_y,\mu}(\ep)$ is the probability amplitude for an electron entering from the left lead at energy $\ep$ and wave-vector $k_y$ to exit in the right lead with energy $\ep+n\hbar\omega$. The expression of the current through the sample is \cite{Moskalets2002,Kohler2005} :
\begin{equation}
I(\mu)=\frac{e}{h}\int_{-\infty}^{\infty}d\ep \sum_{n=-N}^N \left(T_n(\ep) f_L(\ep) - T'_n(\ep) f_R(\ep) \right),
\end{equation}
where $f_L(\ep)$ and $f_R(\ep)$ are the Fermi-Dirac distributions in the left and right leads respectively. The transmission $T_n(\ep)$ is the sum over all $k_y$ channels of the transmission probabilities of electrons from energy $\ep$ in the left lead towards energy  $\ep+n\hbar\omega$ in the right lead :
\begin{equation}
T_n=\sum_{k_y}|t_{n0,k_y}|^2 \, .
\end{equation}

The current flowing in the scatterer is conserved so the scattering matrix is unitary. This implies the relation :
\begin{equation}
\sum_{n=-N}^{n=N} \left(  T_n+R_n \right) = 1 \, ,
\label{unitary}
\end{equation}
where $R_n$ is defined as the sum of the reflection probabilities $|r_{n0,k_y}|^2$ of the transverse modes $k_y$ :
\begin{equation}
R_n=\sum_{k_y}|r_{n0,k_y}|^2 \,.
\end{equation}
We checked that the unitary relation Eq.(\ref{unitary}) is fulfilled in the numerical implementation. We can also notice that $T_n(\ep)=T'_n(\ep)$, $T'_n(\ep)$ being the transmission probability defined similarly as $T_n(\ep)$, but for the reversed scattering process (incident electron going from the right lead and transmitted to the left lead). The equality $T_n(\ep)=T'_n(\ep)$ ensures that there is no pumped current for zero bias. Therefore, at zero temperature, for small bias so that the transmission coefficient varies weakly, we obtain the conductance formula :
\begin{equation}
G(\mu)=\frac{\partial I}{\partial V}=G_0 \sum_{n=-N}^NT_n(\mu),
\label{eq:conductance}
\end{equation}
where $G_0=\frac{4e^2}{h}$, where the factor $4$ accounts for valley and spin degeneracy.

\section{Conductance-chemical potential curves}

\label{sec:doped}
The main goal of this paper is to evaluate the two-terminal conductance $G$ of graphene-based transistors as a function of the chemical potential $\mu$, radiation strength $\beta$, frequency $\omega$, and geometrical parameters $L$ and $W$ (Fig. \ref{fig:transistor}). We consider the ballistic regime relevant for currently achievable high-mobility samples \cite{Calado2014,Banszerus2016,Wang2013One-Dimensional}. 
In this section, the dependence of the conductance upon the chemical potential is discussed for various irradiation strengths $\beta$. After a short review of the non irradiated case ($\beta=0$) \cite{Beenakker2006}, the results are presented by increasing the driving parameter $\beta$ from low to strong driving. The main features consist in strong suppressions of the conductance in wide ranges of chemical potential, especially around $\mu=\pm \hbar \omega/2$. The suppression of the conductance originates from photo-induced gaps in the quasi-energy spectrum which lead to evanescent states.  A simple phenomenological model is introduced which accounts for the residual conductance around  $\mu=\pm \hbar \omega/2$. A crossover between 2D transport through bulk evanescent states and 1D edge transport is predicted, and shown to depend on the geometrical parameters $L$ and $W$ of the graphene ribbon. Specific features at $\mu=0$ (undoped graphene) are discussed more thoroughly in the next section \ref{sec:undoped}.

\subsection{Non-irradiated ribbon $\beta=0$}

\label{sec:non_irrad}

The conductance of a graphene ribbon contacted by two heavily doped leads as a function of the chemical potential has been studied by Tworzydlo {\it et al.}\cite{Beenakker2006}. The system acts as an electronic Fabry-Perot interferometer whose transmission coefficient reads:
\begin{equation}
T^{(p)}_{k_y}=\frac{k_x^2}{k_x^2\cos^2(k_xL)+\left(\frac{\mu}{\hbar v}\right)^2\sin^2(k_xL)},
\end{equation}
for propagating modes with transverse wave-vector $k_{y}=2\pi n_y/W$. The longitudinal wave-vector, which is given by
\begin{equation}
k_x=\pm\sqrt{\left(\frac{\mu}{\hbar v}\right)^2-k_y^2},
\end{equation}
is real for low enough incident angles: $\hbar vk_y < \mu$. 

For larger transverse wave-vector ($\hbar vk_y > \mu$), the longitudinal wave-vector $k_x$ becomes imaginary:
\begin{equation}
k_x=\pm i \kappa_x = \pm i \sqrt{k_y^2 - \left(\frac{\mu}{\hbar v}\right)^2},
\end{equation}
and the transmission is given by:
\begin{equation}
T^{(e)}_{k_y}=\frac{\kappa_x^2}{\kappa_x^2\cosh^2(\kappa_xL)+\left(\frac{\mu}{\hbar v}\right)^2\sinh^2(\kappa_xL)}.
\end{equation}
Since the conductance of the ribbon is obtained by summing over all $k_y$ modes, the transport originates from two contributions : the evanescent modes and the propagating ones, such that the total conductance reads:
\begin{align}
G(\mu)&=G_0\sum_{n_y}T_{k_y}=\frac{G_0W}{2\pi}\int_{-k_\infty}^{k_\infty}dk_yT_{k_y} \\
&=\frac{G_0W}{\pi}\left[\int_0^{\frac{\mu}{\hbar v}}dk_yT^{(p)}_{k_y}+\int_{\frac{\mu}{\hbar v}}^{k_\infty}dk_yT^{(e)}_{k_y}\right] \, ,
\end{align}
where $k_\infty$ is the (very large) Fermi wave vector in the leads. Note that the discrete sum has been replaced by an integral over $k_y$ since large width $W$ is assumed. There is a competition between two contributions : for small $\mu$, the transport will be dominated by evanescent modes, whereas the propagating ones dominate for large $\mu$.

At the Dirac point ($\mu=0$), the conductance originates only from evanescent modes with transmission probability :
\begin{equation}
T^{(e)}_{k_y}=\frac{1}{\cosh^2(k_{y}L)},
\label{eq:trans_evan}
\end{equation}
leading to the minimal conductance:
\begin{align}
G(\mu=0)=\frac{G_0}{\pi}\frac{W}{L}\int^{k_\infty L}_{0}\frac{dk_yL}{\cosh^2(k_yL)}=\frac{G_0}{\pi}\frac{W}{L} \, .
\label{eq:non_irrad}
\end{align}
In this large $W$ regime, the conductance at the Dirac point is proportional to $W$ and $1/L$. It is therefore convenient to normalize the conductance by the ratio $W/L$, as in Fig. \ref{fig:condV01}, green curves, which corresponds to the conductivity $\sigma = GL/W$ of the rectangular ribbon. 

\subsection{Weak driving $\beta=0.1$}

\begin{figure}[h!]
	\includegraphics[width=8cm]{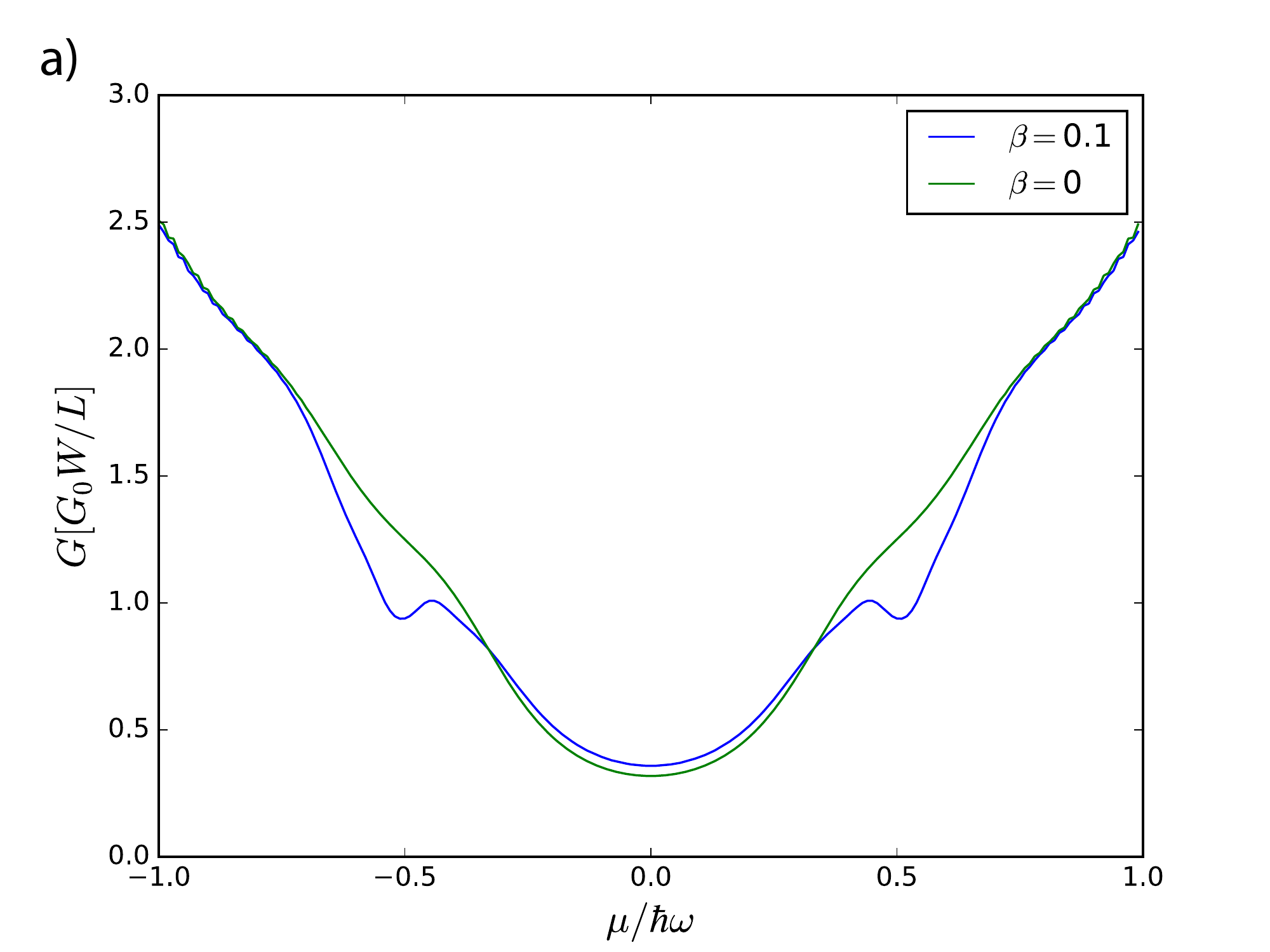}
	\includegraphics[width=8cm]{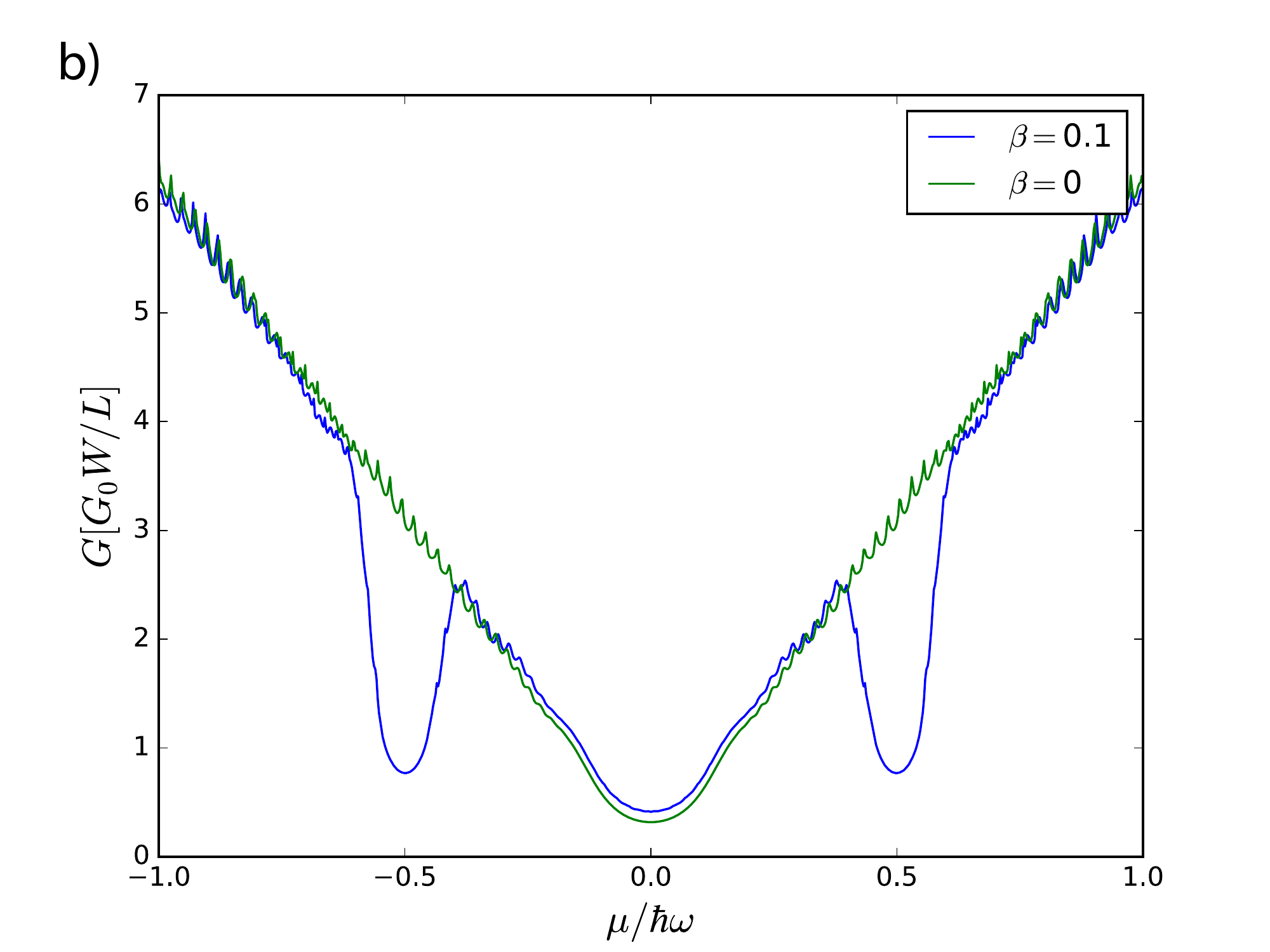}
	\includegraphics[width=8cm]{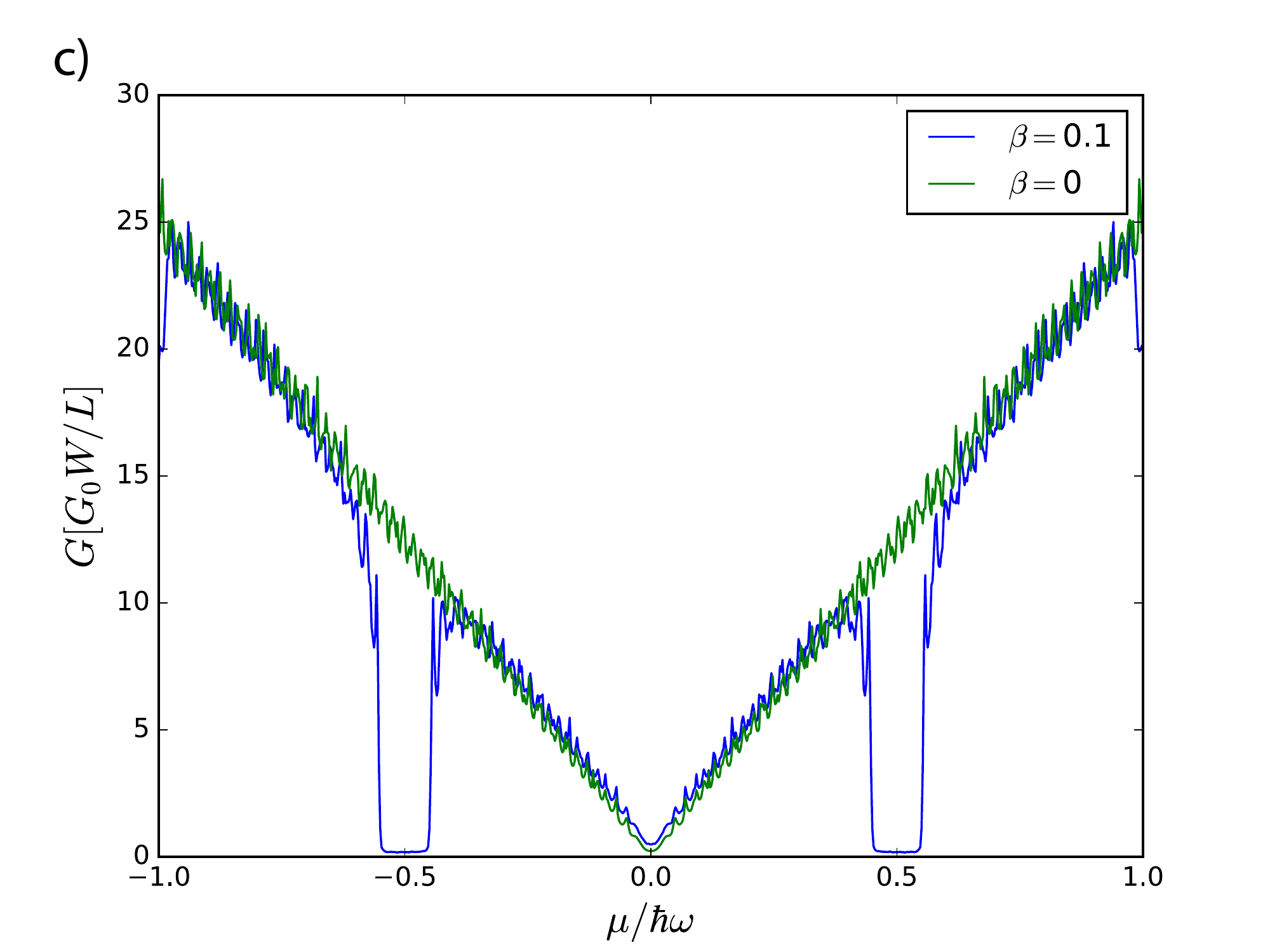}
	\caption{Conductance of a rectangular graphene ribbon as a function of chemical potential for the non-irradiated case ($\beta=0$ in green) and  for $\beta=0.1$ in blue ($N=2$) for a width $W=250 \, l_\omega$ and various lengths : a) $L=10 \, l_\omega=\xi_1$, b) $L=25 \, l_\omega = 2.5 \, \xi_1$ and c) $L=100 \, l_\omega = 10 \, \xi_1$. For the irradiated case, some dips develop around filling $\mu= \pm\hbar \omega/2$ which corresponds to the gaps of order $m=1$ at $k=\pm\omega/2v$ in Fig. \ref{fig:spectrum}.}
	\label{fig:condV01}
\end{figure}

In presence of electromagnetic radiation, the conductance (blue curves in Fig. \ref{fig:condV01}) exhibits broad dips around chemical potentials $\mu=\pm\hbar\omega/2$. These dips mainly correspond to gaps centered at quasi-energy $\ep =\pm  \hbar\omega/2$, and located near $k=\pm \omega/2v$, in the quasi-energy dispersion relation, see Fig. \ref{fig:spectrum}.a). These gaps originate from the one-photon resonance between the valence and the conduction band. The electromagnetic coupling leads to an avoided crossing and the opening of a gap $\Delta_1$, associated to a typical decay length $\xi_1=\hbar v/\Delta_1$. Besides this main gap, the quasi-energy spectrum Fig. \ref{fig:spectrum}.a) also contains a set of very tiny gaps located at higher wave-vectors, around $k=\pm3\omega/2v, \pm5\omega/2v...$, and also all nested around $\pm \hbar \omega/2$. However, for weak driving, the weight of the wave function on these states is negligible, and therefore their contribution to the conductance turns out to be far smaller (than the $\Delta_1$ contribution) although their decay lengths are much larger (nearly propagating states).

\begin{center}
	\begin{table}[h]
		\begin{tabular}{c|c|c|c|c|}
			\cline{2-5}
			\multicolumn{1}{c|}{} & \multicolumn{2}{ c| }{Gaps at $\ep=0$} & \multicolumn{2}{ c| }{Gaps at $\ep=\hbar\omega/2$} \\ \cline{2-5}
			& $m=0$ & $m=2$ & $m=1$ & $m=3$ \\ \cline{1-5}
			\multicolumn{1}{|c|}{$\Delta_m/\hbar\omega$} & 0.02 & 0.01 & 0.1 & 0.0005     \\ \cline{1-5}
			\multicolumn{1}{|c|}{$\xi_m/l_\omega$} & 50 & 100 & 10 & 2000     \\ \cline{1-5}
		\end{tabular}
		\caption{Table of the gap sizes and the characteristic length of the corresponding evanescent states for a driving strength of $\beta=0.1$.}
		\label{ta:gaps01}
	\end{table}
\end{center} 

The conductance at $\mu=\pm\hbar\omega/2$ is therefore controlled by the ratio between the sample length $L$ and the decay length $\xi_1$ of the evanescent state associated to the gap $\Delta_1$, typically $\Delta_1 \simeq 0.1 \, \hbar \omega$ and $\xi_1 = 10 \, l_\omega$ for $\beta=0.1$ according to Eq.(\ref{eq:gapm}) (see also table \ref{ta:gaps01}). Indeed the conductance dips become more pronounced and deeper as the length $L$ increases. For $L=10 \, l_\omega=\xi_1$, the conductance is significantly suppressed (about 25 per cent less than the non irradiated sample conductance) around $\mu=\pm\hbar\omega/2$ [Fig. \ref{fig:condV01}.a)]. For $L=25 \, l_\omega=2.5 \, \xi_1$, the conductance is strongly reduced, roughly by a factor 4, with respect to the non-irradiated value [Fig. \ref{fig:condV01}.b)]. Finally when the length exceeds the penetration length by an order of magnitude, namely for $L=100 \,  l_\omega=10  \, \xi_1$, the conductance dips are sharp and well defined [Fig. \ref{fig:condV01}.c)]. Besides, the width of the dips (which is well-defined only in Fig. \ref{fig:condV01}.c)) corresponds to the value of the gap $\Delta_1$ : $\Delta_1=0.1 \,  \hbar\omega$ (see table \ref{ta:gaps01}). The remaining conductance originates from the states in the gap $\Delta_3$ that have a very long characteristic length. 

Away from these dips, the conductance of the irradiated ribbon follows approximately the non-irradiated one. Here, for $\beta=0.1$, one can notice that the gaps around $\ep=0$ are very small so their characteristic length are much larger than at $\pm\hbar\omega/2$ (Table \ref{ta:gaps01}), which means that even for $L=100 \,  l_\omega$, the deviation from the non-irradiated curve keeps rather small. The conductance near the Dirac point, $\mu=0$, is discussed in more detail in Section \ref{sec:undoped}. We conclude that for weak driving, the effect of the driving on the conductance is mainly observable around $\mu=\pm\hbar\omega/2$. 

\medskip

\subsection{Bulk and edge conductance versus length $L$ for $\beta=0.1$ near $\mu=\pm\hbar\omega/2$}

\label{sec:condL_mu05}

\begin{figure}[t!]
	\includegraphics[width=8cm]{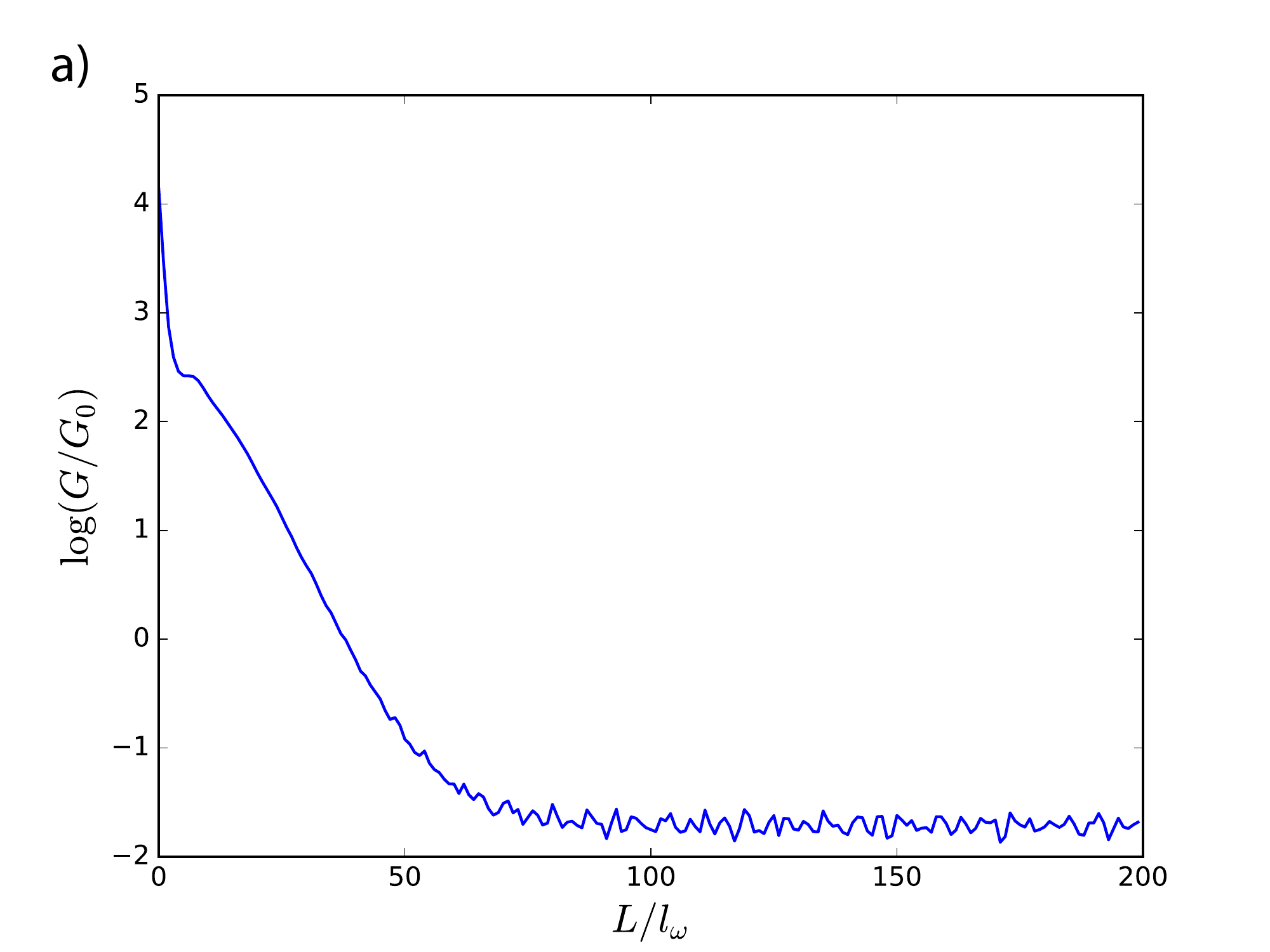}
	\includegraphics[width=8cm]{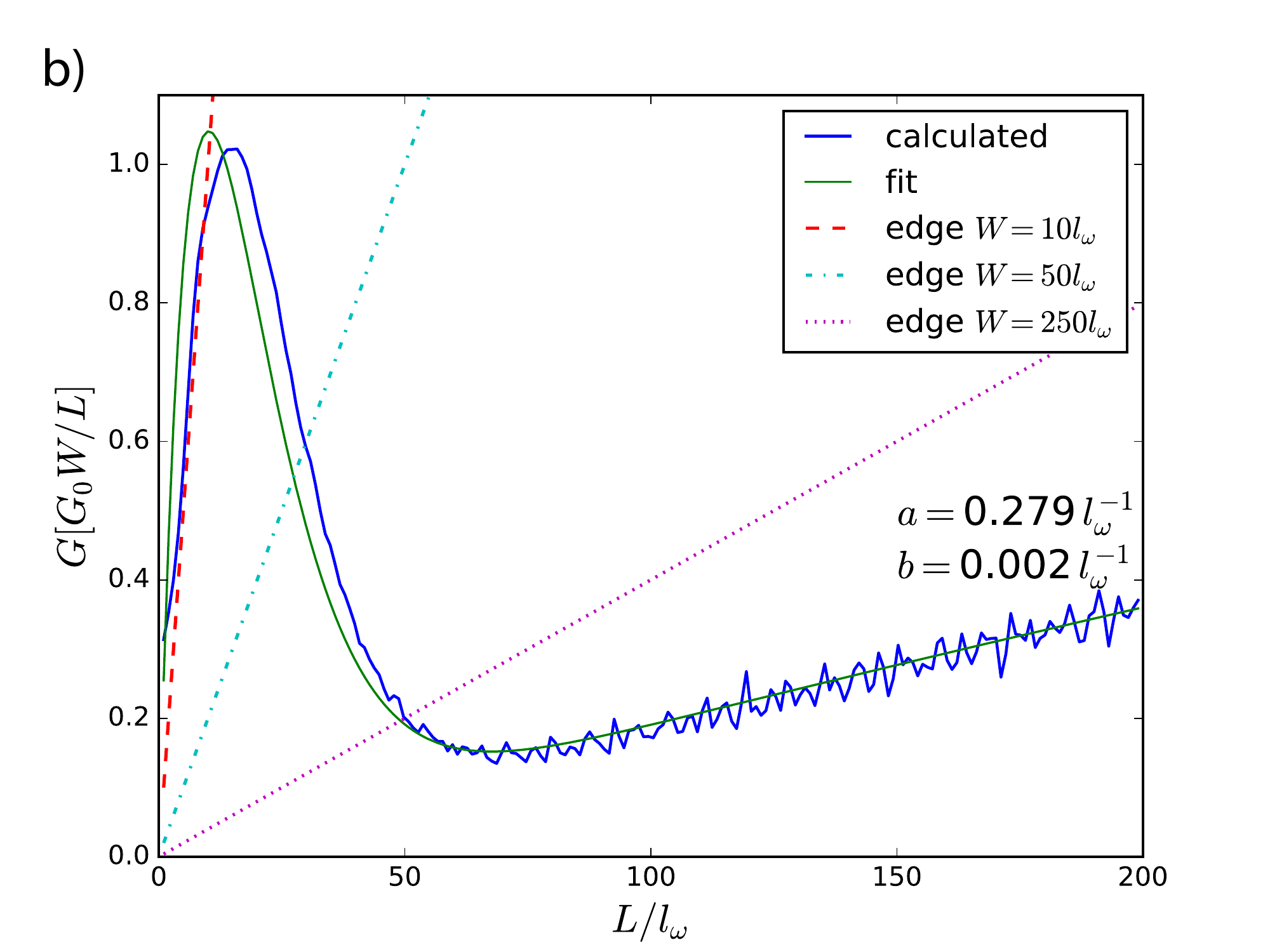}
	\caption{Conductance of a ribbon as a function of the length $L$ of the ribbon for driving strength $\beta=0.1$ and chemical potential $\mu=\hbar\omega/2$ ($N=3$) using two different representation : a) logarithm of the conductance for $W=100l_\omega$, and b) conductance normalized by the ratio $W/L$ (independent of the ribbon width $W$). The green smooth curve corresponds to the fit in Eq. (\ref{eq:fit}), and the resulting fitting parameters are $a$ and $b$. For length  $L\lesssim70l_\omega$, the conductance is carried mainly by the modes with characteristic length $\xi_1$ and for larger length, the modes with length $\xi_3$ dominate.}
	\label{fig:condL_V01}
\end{figure}

\textit{Bulk states} : Inside the dips at $\mu=\pm\hbar\omega/2$, the current is mainly carried by evanescent modes, thus, the conductance of the sample is expected to decrease exponentially with the length. At short length, the conductance in the gap $\ep=\hbar\omega/2$ is dominated by the evanescent modes coming from the gap $\Delta_1$, but for larger length, the modes originating from the gap $\Delta_3$ will be dominant. Therefore, it is possible to express the conductivity in the gap $\ep=\hbar\omega/2$ with a simple model taking into account this interplay between states at gaps $\Delta_1$ and $\Delta_3$. First, we have checked that the conductance is proportional to $W$, and thus our ansatz reads:  
\begin{equation}
	G_{bulk} =G_0 W \left(a e^{-L/\xi_1}+b e^{-L/\xi_3}\right) ,
\label{eq:fit0}
\end{equation}
where $\xi_1$ and $\xi_3$ are given by Table \ref{ta:gaps01}, $a$ and $b$ being the only fitting parameters of our model that depend only on $\beta$. The latter parameters (expressed in units of $l_\omega^{-1}$) quantify the relative importance of the evanescent states in transport. Finally, using the rescaling by $G_0 W/L$, we have plotted in Fig. \ref{fig:condL_V01}.b) the following ratio: 
\begin{equation}
	\frac{G}{G_0} \frac{L}{W}=L \left(a e^{-L/\xi_1}+b e^{-L/\xi_3}\right) \,,
\label{eq:fit}
\end{equation}
and extracted the value of $a$ and $b$. The peak with maximum around $L=10 \, l_\omega$ corresponds to the current carried by the evanescent modes carried mainly by the evanescent states originating from the first order side-band ($\xi_1=10l_\omega$). The curve reaches a minimum around $L=70 \, l_\omega$ and then increases again as the states coming from the third order side-band become the dominant source of current. The parameter $b$ is two orders of magnitude smaller than $a$, which indicates that for $\beta=0.1$, the weight of the wave function on the evanescent states corresponding to the gap $\Delta_3$ is very weak. This fact corroborates the approximation we made by considering only the evanescent states originating from the gaps $\Delta_1$ and $\Delta_3$.

\medskip

\begin{figure}[h!]
	\includegraphics[width=8cm]{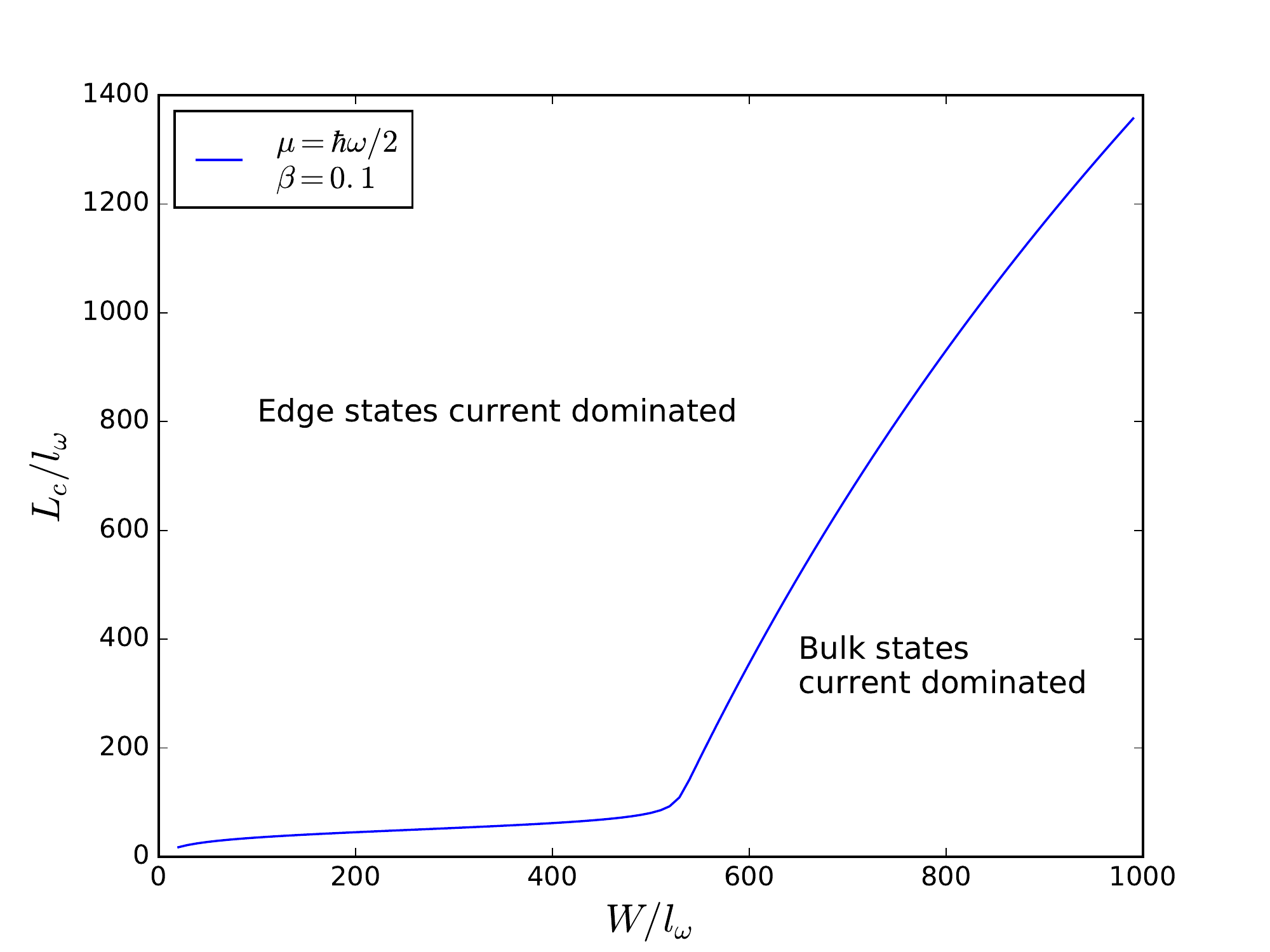}
	\caption{Critical length $L_c$ where bulk and edge have the same contribution to the current at $\mu=\hbar\omega/2$ for a driving strength $\beta=0.1$. For lengths longer than $L_c$, the current is carried mainly by edge states whereas for shorter lengths, the current is dominated by the evanescent bulk states.}
	\label{fig:phase_diagV01}
\end{figure}

\textit{Edge states} : So far only transport through 2D bulk states (propagating or evanescent) has been evaluated and discussed. One can evaluate the edge state contribution and compare it with the bulk contribution. The dotted lines corresponds to the edge state contribution. There is one edge state per valley that contributes to the dc conductance\cite{Usaj2014,Foa2014,Perez-Piskunow2015}, taking into account for valley and edge degeneracy, the maximal conductance is equal to :
\begin{equation}
G_{edge}(\mu=\hbar\omega/2)=\frac{4e^2}{h} \, .
\end{equation}
We said ``maximal" conductance because the irradiation may reduce the conductance of the edge states. This effect has been studied by Aaron \textit{et al.}\cite{Aaron2015} using the Bernevig-Hugues-Zhang model\cite{Bernevig2006}. They found that the conductance is reduced by a Bessel factor with argument $\beta$, in the same fashion as in the photon-assisted transport problem. In this paper, as we consider weak driving, we will not take into account this reduced edge state conductance. Inter-edge scattering is neglected as only large sample widths $W$ are considered. We have plotted the ratio $G_{edge}L/G_0W$ for the edge states for different widths of the sample (Fig. \ref{fig:condL_V01}). This demonstrates that the conductance is dominated by edge states only for long enough samples, namely when $L$ exceeds the typical decay length of the bulk evanescent states. This crossover between 2D transport by evanescent states and 1D edge transport occurs at a typical length $L_c$ which increases with the ribbon width $W$. Using the fit function (\ref{eq:fit0}) with the parameters $a$ and $b$, it is possible to plot $L_c$ as a function of $W$ using the relation :
\begin{equation}
	G_{bulk}(W,L_c)=G_{edge}\, .
\end{equation}

The resulting $L_c(W)$ curve is plotted in Fig. \ref{fig:phase_diagV01}. This ``phase diagram" allows us to see the competition between bulk and edge current which depends on the shape of the sample and the frequency of irradiation (through $l_\omega$). This curve shows that bulk transport dominates for short and wide graphene ribbons which can be understood qualitatively. Quantitatively, it gives a criterium needed to separate bulk and edge conductance in a particular experiment. There is a clear break in the curve around $W=500l_\omega$, where the edge conductance becomes smaller than the conductance carried by bulk evanescent states originating from the $\Delta_3$ gap.

\subsection{Electromagnetic driving $\beta=0.3$}

\begin{figure}[h!]
	\includegraphics[width=8cm]{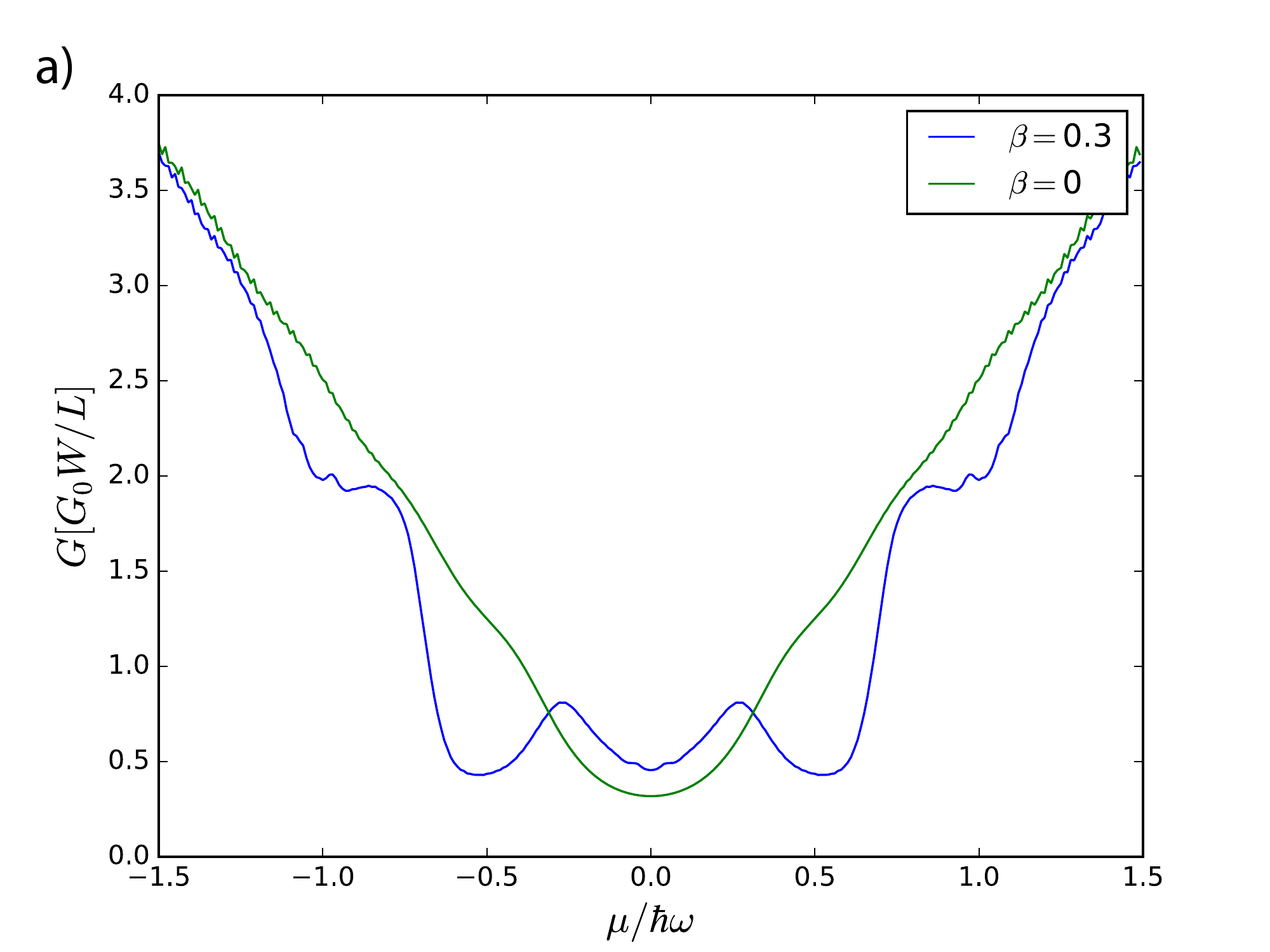}
	\includegraphics[width=8cm]{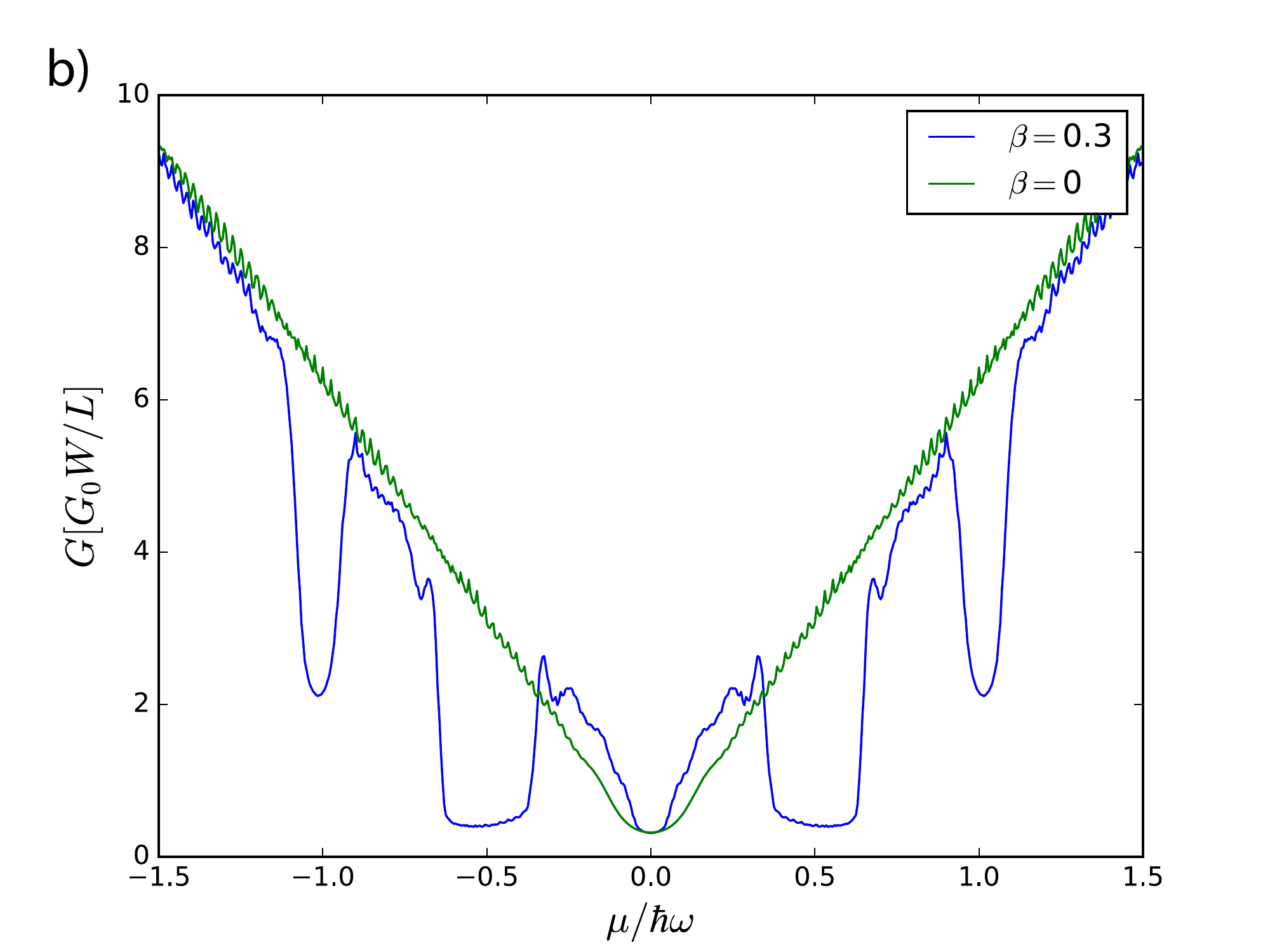}
	\includegraphics[width=8cm]{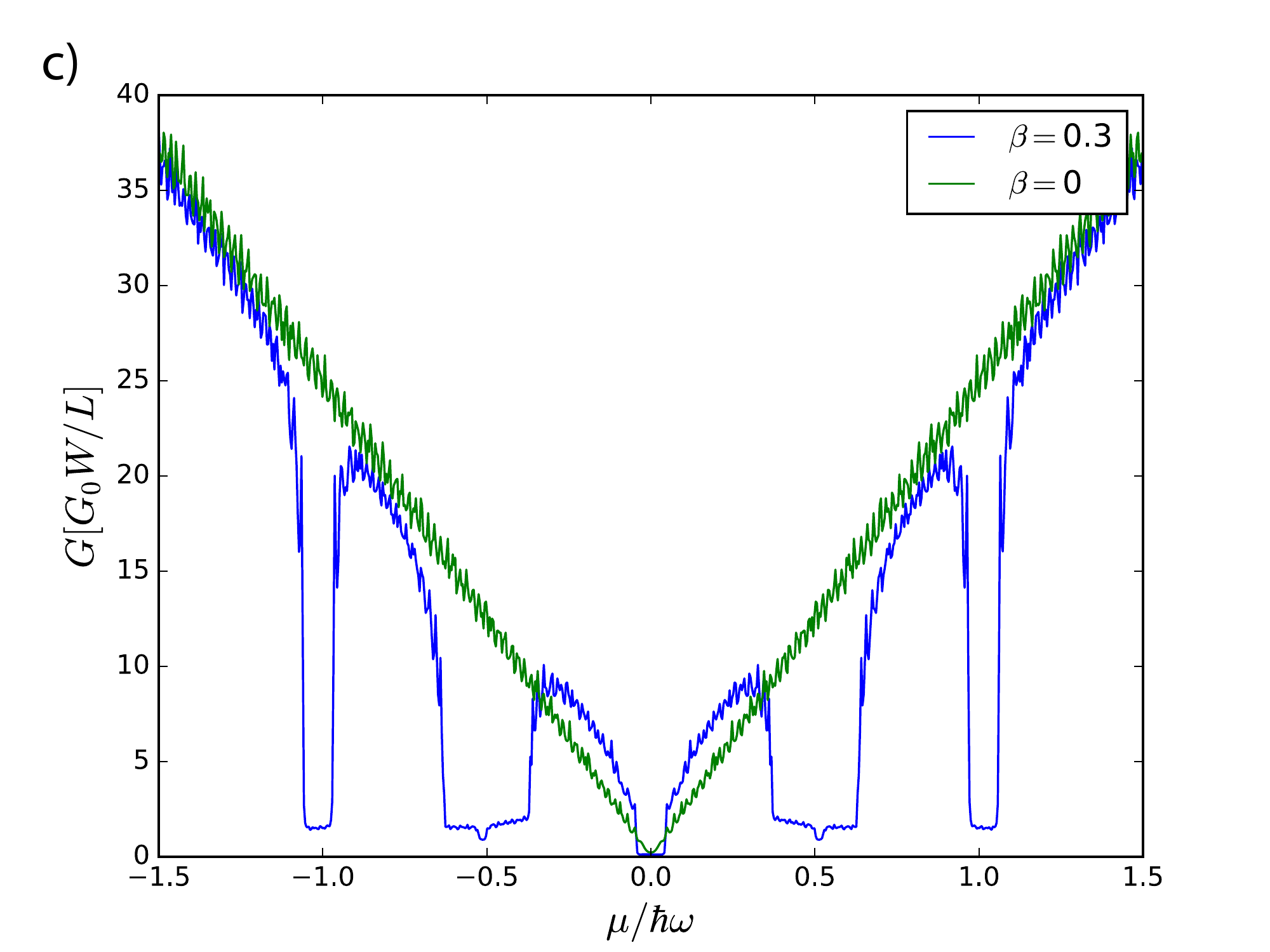}
	\caption{Conductance of an irradiated graphene ribbon as a function of chemical potential for the non-irradiated case ($\beta=0$, green curve) and  for $\beta=0.3$ ($N=2$, blue curve) for a width $W=250l_\omega$ and a length : a) $L=10l_\omega$, b) $L=25l_\omega$ and c) $L=100l_\omega$. In the irradiated case, some dips develop around filling $\mu=0,\pm\hbar \omega/2$ and $\pm\hbar \omega$ which corresponds to the gaps of order $m=1$ at $k=\pm\omega/2v$ and the gaps in fig. \ref{fig:spectrum}.}
	\label{fig:condV03}
\end{figure}

For $\beta=0.3$, the scenario is similar to the case with $\beta=0.1$, except that some dips develop at $\mu=0$ and $\pm\hbar\omega$ in addition to the dips at $\pm\hbar\omega/2$. Those dips correspond to the second order processes in $\beta$ which are not negligible anymore. Table \ref{ta:gaps03} shows the gaps sizes and the length of the corresponding evanescent states. As the length increases, the dips develop and are better defined.
 
 \begin{center}
 	\begin{table}[h]
 		\begin{tabular}{c|c|c|c|c|c|}
 			\cline{2-6}
 			\multicolumn{1}{c|}{} & \multicolumn{3}{ c| }{Gaps at $\ep=0$} & \multicolumn{2}{ c| }{Gaps at $\ep=\hbar\omega/2$} \\ \cline{2-6}
 			& $m=0$ & $m=2$ & $m=4$ & $m=1$ & $m=3$ \\ \cline{1-6}
 			\multicolumn{1}{|c|}{$\Delta_m/\hbar\omega$} & $0.16$ & 0.08 & 0.0013 & 0.3 & 0.0135     \\ \cline{1-6}
 			\multicolumn{1}{|c|}{$\xi_m/l_\omega$} & 6.25 & 12.5 & 761 & 3.33 & 74.1     \\ \cline{1-6}
 		\end{tabular}
 		\caption{Table of the gaps size and the characteristic length of the corresponding evanescent states for a driving strength of $\beta=0.3$.}
 		\label{ta:gaps03}
 	\end{table}
 \end{center}

Each gap is larger at $\beta=0.3$ than its value at $\beta=0.1$ case (Fig. \ref{fig:gaps0}). Hence the corresponding evanescent state decay lengths have all decreased, thereby making the residual conductance larger, provided $L$ is kept constant. However, the weight of the wave-function on the higher order side-bands has increased. In the gaps at $\mu=\pm\hbar\omega/2$, we can see in Fig. \ref{fig:condV03}.c) that the conductance is not zero anymore because the third order side-bands now carry some current. We can also see the presence of the gap $\Delta_3$ at $\mu=\pm\hbar\omega/2$ inside the gap $\Delta_1$ because there is a small dip in the middle inside the larger dip. Nevertheless, the gap at $\mu=0$ is perfectly defined for $L=100l_\omega$ because the current carried by the fourth order side-bands is zero. In the gaps at $\mu=\pm\hbar\omega$, the current is not zero because the density of states is much higher than in the gap at $\mu=0$.
 
 \medskip

\section{Undoped graphene}

\label{sec:undoped}

In this section, we analyse more thoroughly the contribution of bulk states to the conductance when the chemical potential is equal to zero. We discuss the evolution of the conductance at the Dirac point as function of the irradiation strength $\beta$, and as a function of length $L$, comparing it with the edge contribution. We compare our results with those of Gu {\it et al.} \cite{Gu2011}. 


\subsection{Minimal conductivity vs driving strength}

At the Dirac point, for $\beta=0$, the current is carried by evanescent states that belongs to the elastic channel. Due to the semi-metallic nature of graphene, their decay length is infinite for $k_y=0$ and decreases as $|k_y|$ increases (Eq. \ref{eq:trans_evan}). For $\beta\neq0$, the Floquet bands are coupled, so the inelastic channels with $n\neq0$ are opened and the whole set of evanescent states located at wave-vector $k_x^m=\pm m\omega/2v$ will contribute to the current. The evanescent states in the gap at $k=0$ that originate from the elastic channel $n=0$ are now gapped, and their decay length is no longer infinite.

\begin{figure}[h!]
\includegraphics[width=9cm]{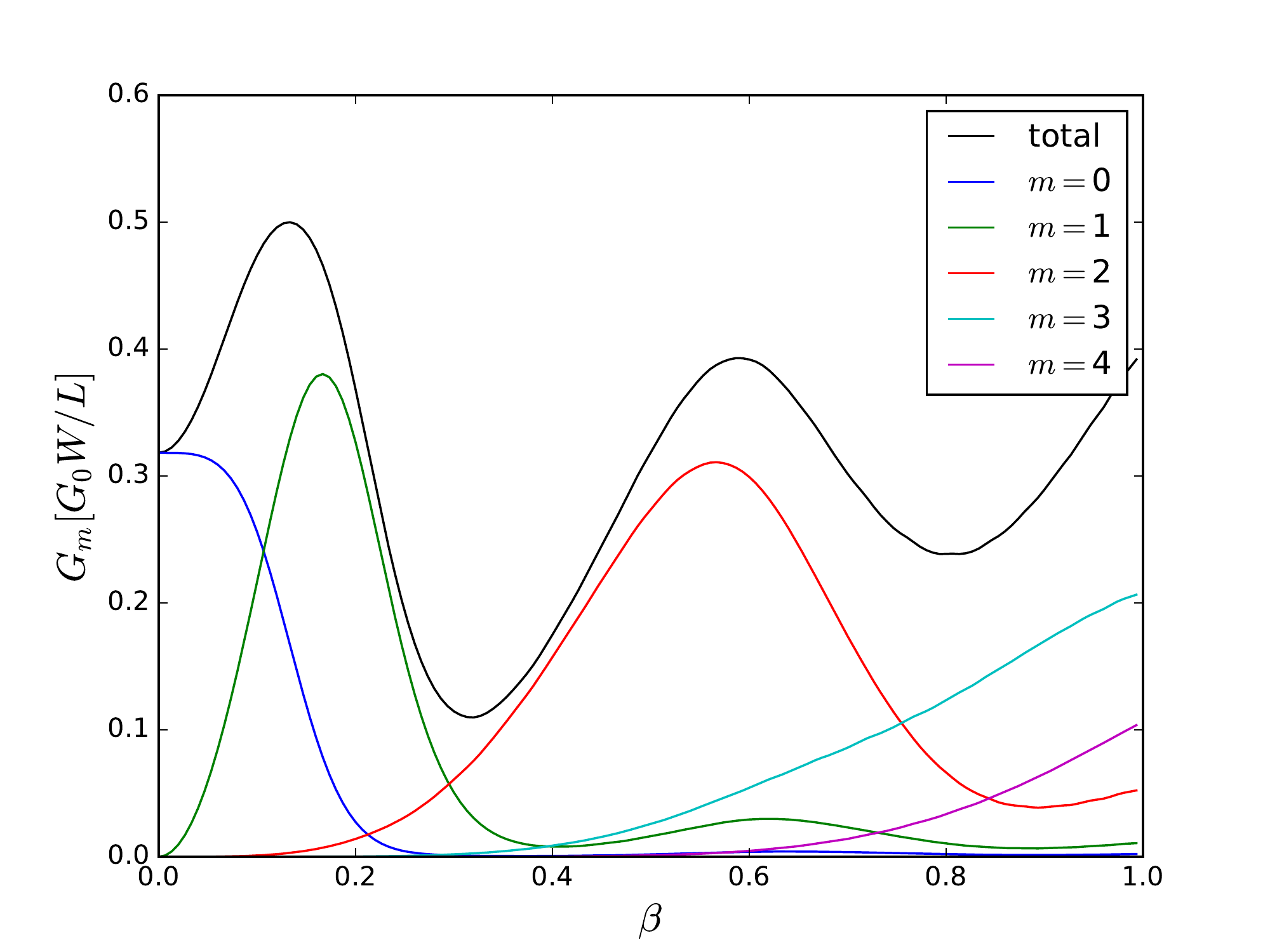}
\caption{Conductance $G_{m}$ at the Dirac point carried by electrons having emitted or absorbed $m$ photons in function of the driving strength for a length  $L=50l_\omega$ and $N=5$ Floquet replicas. $G_{m}$ is defined by Eq. (\ref{eq:cond_m}) in the text. The competition between the decreasing characteristic length of the evanescent states and the increasing weight of the wave function over replicas with higher number of photons creates these oscillations.}
\label{fig:cond_V}
\end{figure}

The characteristic length of the evanescent states in the gap of order $m$ is given by Eqs. (\ref{eq:length_evan1}) and (\ref{eq:length_evan2}). When $\beta$ increases, their decay length decreases and therefore the current carried by the evanescent state with number of photon $n$ should decrease. However, with increasing $\beta$, the amplitude of the wave function over states with higher number of photons $n$ inside the irradiated region increases. There is a competition between these two effects that creates oscillations of the conductance at the Dirac point when varying $\beta$ (Fig.\ref{fig:cond_V}).

To understand this competition, it is useful to plot the conductivity $G_{n}$ carried by each channel $n$, where $n$ corresponds to the number of photons absorbed or emitted, defined as : 
\begin{equation}
G_n=\frac{4e^2}{h}(T_n+T_{-n})
\label{eq:cond_m}
\end{equation}
evaluated at $\mu=0$. We normalize this quantity by $W/L$ in order to compare our results to the non-irradiated case. We realize that the shape of this curve is independent of W which means that the conductance G is a linear function of W. This is consistent with the fact that the density of states is proportional to $1/W$. However this curve depends on $L$ because $L$ has to be compared to the characteristic length $\xi$ of the evanescent states (which are functions of $\beta$). The effect of varying L on this curve is to compress or expand the curve along the $\beta$ axis. In the next section, we study the evolution of the conductance as a function of L for a fixed $\beta$.

For $\beta=0$, we recover the non-irradiated result minimal conductance:
\begin{equation}
G(\mu=0,\beta=0)=\frac{G_0}{\pi}\frac{W}{L} \, .
 \end{equation}

For $\beta<0.15$, the two main contributions come from the channels $m=0$ and $m=1$ (Fig. \ref{fig:cond_V}). As expected, the current carried by the channel $n=0$ decreases with $\beta$ because : $i$) the gap $\Delta_0$ increases and $ii$) the weight of the wave function on this channel decreases. We now turn to the current carried by the channel $n=1$ : for $\beta=0$ it is zero because there is no irradiation, thus it necessarily increases with $\beta$ because we open this channel. We can see in Fig. \ref{fig:cond_V} that the conductance $G_1$ increases faster than $G_0$ decreases, therefore the total conductance increases.

At $\beta=0.15$, the channel $m=2$ starts to contribute although the main part of the current is being carried by channel $m=1$. Then for $\beta>0.2$, the conductance $G_1$ of this channel ($m=1$) starts to decrease while the conductance $G_2$ increases with $\beta$. Around $\beta=0.3$, the total conductance $G$ reaches a minimum, while contributions from the $n=1$ and $n=2$ channels are almost equal. The same scenario repeats with increasing $\beta$.

\subsection{Minimal conductivity vs length}

\begin{figure}[h!]
	\includegraphics[width=8cm]{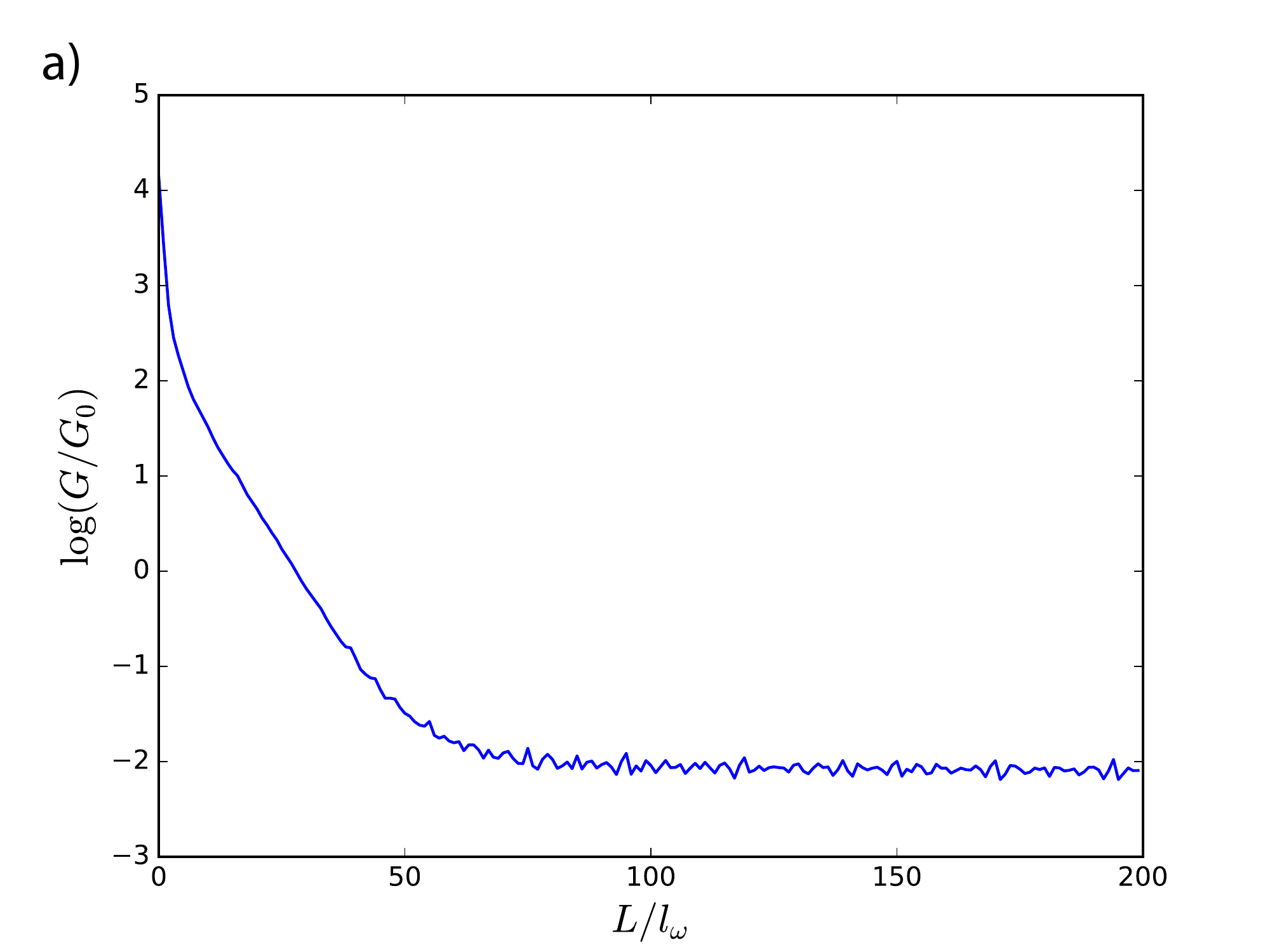}
	\includegraphics[width=8cm]{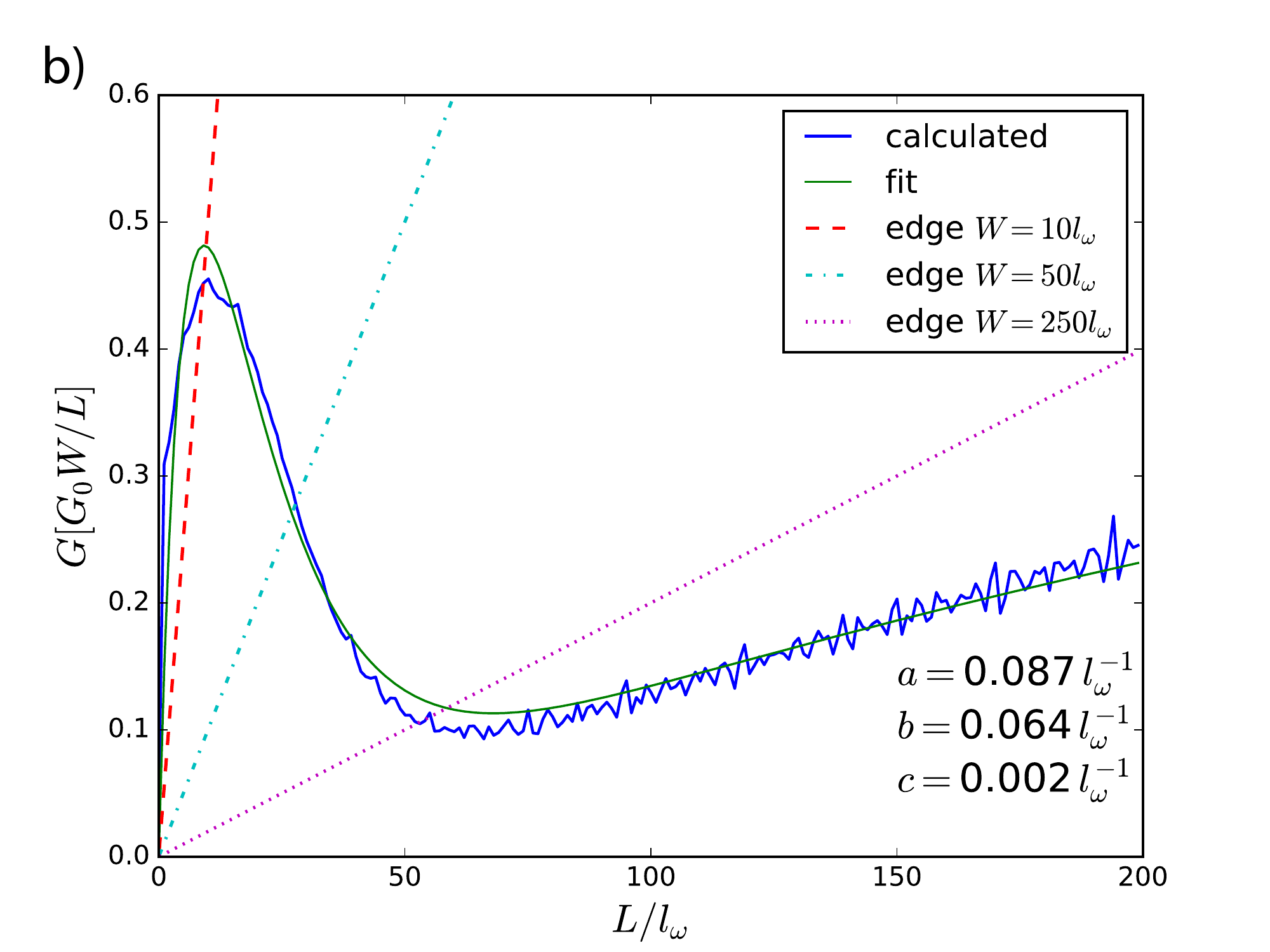}
	\caption{Conductance of a ribbon as a function of the length $L$ of the ribbon for driving strength $\beta=0.3$ and chemical potential $\mu=0$ ($N=3$) using two different representation : a) logarithm of the conductance for $W=100l_\omega$, and b) conductance normalized by the ratio $W/L$ (this curve is independent of the ribbon width $W$). The smooth green curve corresponds to the fit in Eq. (\ref{eq:fit_0}), and the resulting fitting parameters are $a$, $b$ and $c$. The dashed lines corresponds to the edge state conductance given by Eq. (\ref{eq:edge_0}), which are linear with respect to $L$ because of their ballistic nature.}
	\label{fig:condL_V03}
\end{figure}

For $\beta=0.3$, we obtain that the conductance can be fitted by a simple model including only 3 evanescent states (associated to 3 nested gaps around $\ep=0$ in Fig. \ref{fig:spectrum}.a) ) :
\begin{equation}
\frac{G}{G_0}\frac{L}{W} =L \left(  ae^{-L/\xi_0}+be^{-L/\xi_2}+ce^{-L/\xi_4}   \right) \, ,
\label{eq:fit_0}
\end{equation}
where the fitting parameters $a$, $b$ and $c$ are shown on the plot. For $\beta=0.3$, the conductance originating from evanescent states that belong to the $\Delta_4$ gap is very weak. We checked that adding the conductance states originating from the gap $\Delta_6$ in the fitting function (\ref{eq:fit_0}) doesn't change the results (for lengths smaller than $800l_\omega$, according to Table \ref{ta:gaps03}). Therefore the scaling of the bulk conductance is quite different than the one obtained by Gu {\it et al.} \cite{Gu2011}. Our model suggests that the bulk conductance is described by Eq. \ref{eq:fit_0}) where only the evanescent states with small $m$ ($m$ is the order of the anti-crossing) are required, whereas in Gu {\it et al.}, the whole set of evanescent states originating from the gaps $\Delta_m$ including those with high $m$ has to be taken into account, resulting in an approximate power law behavior\cite{Gu2011}.

\begin{figure}[h!]
	\includegraphics[width=9cm]{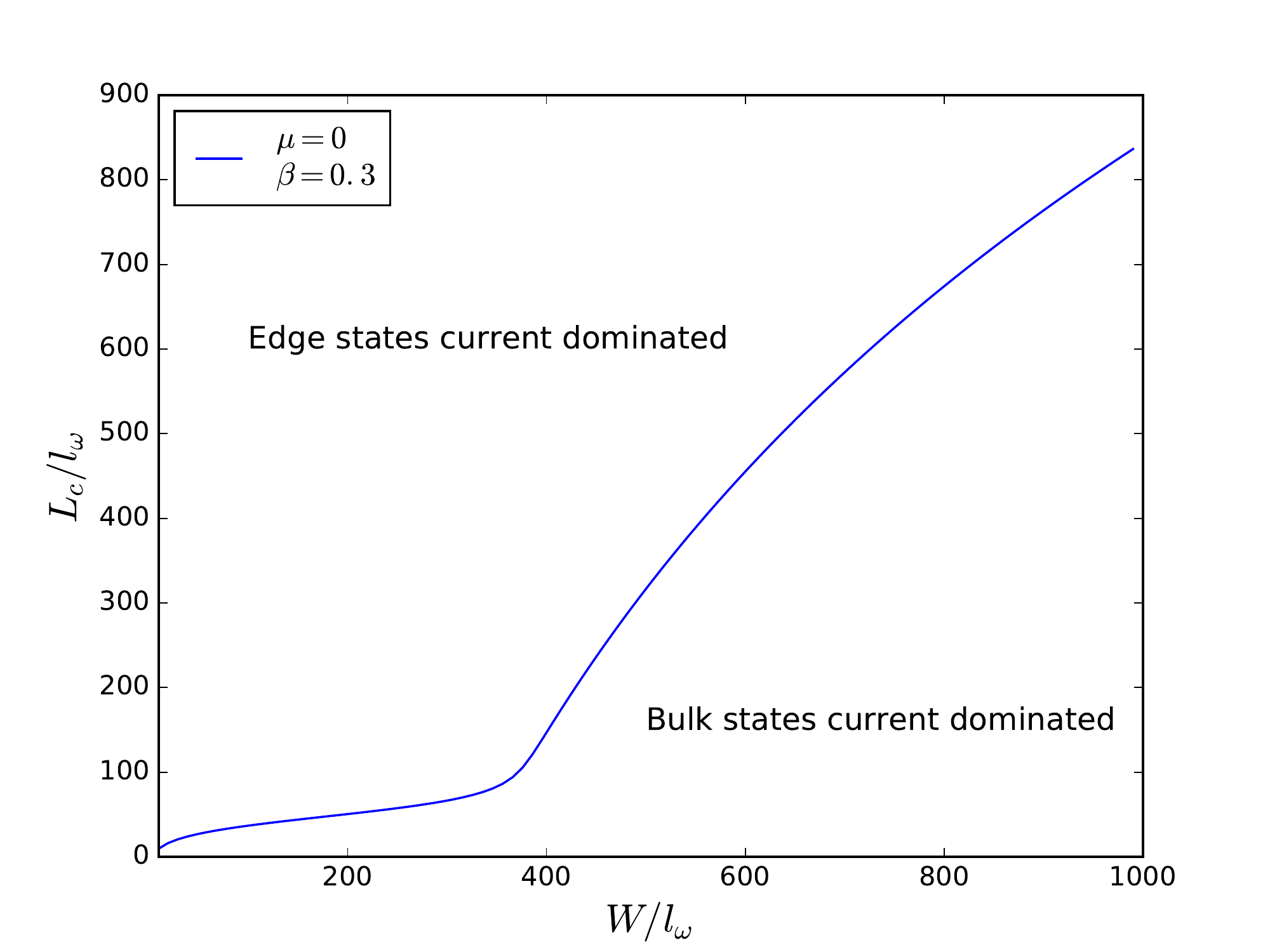}
	\caption{Critical length $L_c$ where bulk and edge have the same contribution to the current at $\mu=0$ for a driving strength $\beta=0.3$. For lengths longer than $L_c$, the current is carried mainly by edge states whereas for shorter lengths, the current is dominated by the evanescent bulk states.}
	\label{fig:phase_diagV03}
\end{figure}

\textit{Edge state conductance :} So far, only the bulk conductance has been evaluated and it is necessary to explicit the conditions upon which edge state contributions dominates. The dotted lines in Fig. \ref{fig:condL_V03} corresponds to the edge state conductance. There is one edge state linking the valleys that contributes to dc conductance\cite{Gu2011,Perez-Piskunow2015,Kitagawa2011,Foa2014}, therefore the maximal edge conductance is equal to :
\begin{equation}
G_{edge}(\mu=0)=\frac{2e^2}{h}=G_0/2.
\label{eq:edge_0}
\end{equation}
This situation is completely identical to the case studied in Sec. \ref{sec:condL_mu05} except for the factor $1/2$ in the edge conductance and the different expression for the bulk conductance (\ref{eq:fit_0}). We don't take into account the non-quantization of the edge states due to the driving\cite{Aaron2015} but we know that the edge states are chiral and robust to disorder, so that their conductance is independent of the size of the sample. We apply the same procedure as in Sec. \ref{sec:condL_mu05} to see the competition between edge and bulk states, and obtain the curve $L_c(W)$ for which the edge and the bulk states conductance is equal (Fig. \ref{fig:phase_diagV03}).


\section{Experimental parameters}

In this section, we discuss the assumptions we made, and the restrictions they put on experimental realizations of such an irradiated graphene gFET. First, we investigate what the restrictions on the system parameters are necessary in order to have a ballistic conductor, meaning that dissipation occurs in the leads. In a second part, we consider the laser characteristics (frequency and power) that are needed to observe the particular effects predicted in this paper.

\subsection{Dissipation}

The Landauer-B\"{u}ttiker formalism implicitly assumes that all the dissipation occurs in the leads. In the irradiated region, the electrons are coherently dressed by one or several quanta of the electromagnetic radiation, and the corresponding excess energy is dissipated in the drain electrode of the transistor. Clearly, this is only valid when the irradiated region length $L$ is short enough. If we consider coupling to a phonon bath, we have to compare the characteristic time for excited electrons to emit phonons to the passage time of the electron through the scattering region. We require
\begin{equation}
L< v\tau_{ph}
\end{equation}
where $\tau_{ph}$ is the average time for an electron to decay into phonons. This gives an upper bound on the sample length, depending on which process dominates at the energy of the photons.
The dominant process for relaxation of excited electrons is the emission of optical phonons of energy 194 or 330 meV\cite{Sun2008}. The characteristic time for this process is $\tau_{opt}\approx1\text{ps}$, which limits the sample length to $L<1$\si{\micro\meter}.
For electrons with smaller energy, the only dissipating channel is the acoustic phonon \cite{Tse2009,Bistritzer2009}, which is a much slower process that takes place on a timescale of the order of a nanosecond\cite{Bistritzer2009}. This channel limits the length to $L<1$ mm.

Finally, the assumption of a ballistic conductor requires the length of the sample to be smaller than the mean free path $l_e$ of scattering of electrons by impurities, which in high-quality graphene is $l_e \approx \,$ 1-10\si{\micro\meter}.

In order to safely neglect the effect of dissipation by scattering of electrons by optical phonons, we then need electron energies $\hbar\omega$ smaller than 200 meV. In this regime, the two remaining dissipation processes that limit the sample size are scattering by impurities and acoustic phonons. Since the acoustic phonon scattering time is very long, the only remaining process restricting the size of the sample for electron energies $\hbar\omega<200$ meV is impurity scattering.
At these typical energies, the photon energy is much smaller than the bandwidth of graphene, so the Dirac equation approximation is valid.

\subsection{Observability}

Accounting for the restriction on the photon energies in the last section, we consider photon energies ranging from 10 to 100 meV.
Experimentally, the typical electron densities that can be reached in graphene are around $5\cdot10^{12}$ \si{\centi\meter\per\second} which corresponds to Fermi energies up to 250 meV. With the electron energies considered, it is experimentally feasible to reach the energy dips at $\pm\hbar\omega/2$.

Photons with energy from 10 to 100 meV corresponds to Terahertz frequencies ranging from 2.5 to 25 THz. This frequency window corresponds to characteristic lengths $l_\omega=6$ nm to 0.6 nm. Using Eq. \ref{eq:electric_field}, the electric field intensity $E_0$ required to reach $\beta=0.1$ range from $10^5$ to $10^7\si{\volt\per\meter}$.

\section{Conclusions}

\label{sec:conclusions}

The two-terminal conductance of a rectangular graphene ribbon irradiated by an electromagnetic wave (frequency $\omega$) has been studied using the Floquet theory for experimentally realizable setups. In the ballistic regime our numerical calculations confirm that the coherent dressing of the original Dirac cone leads to the opening of a set of two non-equivalent photo-induced gaps in the Floquet zone, around quasi-energies $\ep=0$ and $\ep=\hbar\omega/2$. The size of the gaps are found to be in good agreement with RWA estimations\cite{Zhou2011} at low irradiation $\beta$. 
When the chemical potential is tuned inside one of these photo-induced gaps, the conductance of the sample decreases drastically because the current is now carried by evanescent states (rather than propagating states when $\beta=0$). The conductance curve as a function of the chemical follows closely the non-irradiated one except at integer multiples of $\hbar\omega/2$ where broad dips appear. For weak driving, the main effect on the conductance is seen at doping $\mu=\hbar\omega/2$, while for stronger driving additional dips also develop around $\mu=\hbar\omega$. The effect of the irradiation is also effective at the Dirac point where it modulates the value of bulk minimal conductance. The widths of these dips (on the axis of chemical potential) correspond to the sizes of the gaps in the Floquet spectrum, while the depth of these dips (on the conductance axis) depends on the length of the sample. The residual conductance in dips has been fitted by assuming that the current is carried by a few evanescent states characterized by distinct decay lengths. 

Besides the transport by 2D bulk states, electrons may also propagate through photo-induced 1D edge states. We have shown that the edge state contribution to the total conductance can be neglected for short enough samples, and quantitative criteria to be in the bulk transport regime have been given.  

\medskip

\section*{Acknowledgements}

We are grateful to Dario Bercioux and Bogusz Bujnowski for useful comments. JHB was supported by the ERC Starting Grant No.~679722.

\bibliographystyle{apsrev4-1}
\bibliography{bibl}
\end{document}